\documentclass[%
 reprint,
 amsmath,amssymb,
 aps,
]{revtex4-2}
\usepackage{graphicx}
\usepackage{dcolumn}
\usepackage{bm}
\usepackage[colorlinks,linkcolor=blue,citecolor=blue,urlcolor=blue]{hyperref}

\begin{document}

\preprint{APS/123-QED}

\title{Onsager Reciprocal Relation between Anomalous Transverse Coefficients of an Anisotropic Antiferromagnet}

\author{Xiaodong Guo$^{1}$, Xiaokang Li$^{1}$, Zengwei Zhu$^{1,*}$ and Kamran Behnia$^{2,\dagger}$}

\affiliation{$^1$Wuhan National High Magnetic Field Center and School of Physics, Huazhong University of Science and Technology,  Wuhan,  430074, China\\
$^2$Laboratoire de Physique et Etude des Mat\'{e}riaux (CNRS/UPMC),Ecole Sup\'{e}rieure de Physique et de Chimie Industrielles, 10 Rue Vauquelin, 75005 Paris, France\\
}

\date{\today}
\begin{abstract}
Whenever two irreversible processes occur simultaneously, time-reversal symmetry of microscopic dynamics gives rise, on a macroscopic level, to Onsager's reciprocal relations, which impose constraints on the number of independent components of any transport coefficient tensor. Here, we show that in the antiferromagnetic YbMnBi$_2$, which displays a strong temperature-dependent anisotropy, the Onsager's reciprocal relations are strictly satisfied for anomalous electric ($\sigma^A_{ij}$) and anomalous thermoelectric  ($\alpha^A_{ij}$) conductivity tensors. In contradiction with what was recently reported by Pan \textit{et al.} [Nat.Mater. 21, 203 (2022)], we find that $\sigma^A_{ij} (H)= \sigma^A_{ji} (-H)$, and $\alpha^A_{ij} (H)= \alpha^A_{ji} (-H)$. This equality holds in the whole temperature window irrespective of the relative weights of the intrinsic or extrinsic mechanisms. The $\alpha^A_{ij}/\sigma^A_{ij}$ ratio is close to $k_B/e$ at room temperature, but peaks to an unprecedented magnitude of 2.9 $k_B/e$ at $\sim$ 150 K, which may involve nondegenerate carriers of small Fermi surface pockets.

\end{abstract}

\maketitle

Onsager's reciprocal relations are derived by assuming that all microscopic processes are reversible: The future is symmetric to the past \cite{Onsager-1,onsager-2, mazur1954extension, de1954extension, callen1948application, Casimir1945, Kamran2015}. This is called time-reversal symmetry. Given a system brought out of equilibrium by the thermodynamic forces $F_k$, the corresponding fluxes $J_i$ are that, in the coupled transport equation $J_i=\sum_kL_{ik}F_k$, the kinetic coefficients $L_{ik}$ obey the relation $L_{ik}=-L_{ki}$. The time-reversal symmetry is broken when a magnetic field $H$ is applied. This relation states that the kinetic coefficients obey $L_{ik}(H)=L_{ki}(-H)$. This relation, a cornerstone of nonequilibrium statistical physics, is relevant to any case of two irreversible processes occurring simultaneously. 

\begin{figure}[ht]
\setlength {\abovecaptionskip} {0.1cm}
\setlength {\belowcaptionskip} {-0.6cm}
\includegraphics[width=8.5cm]{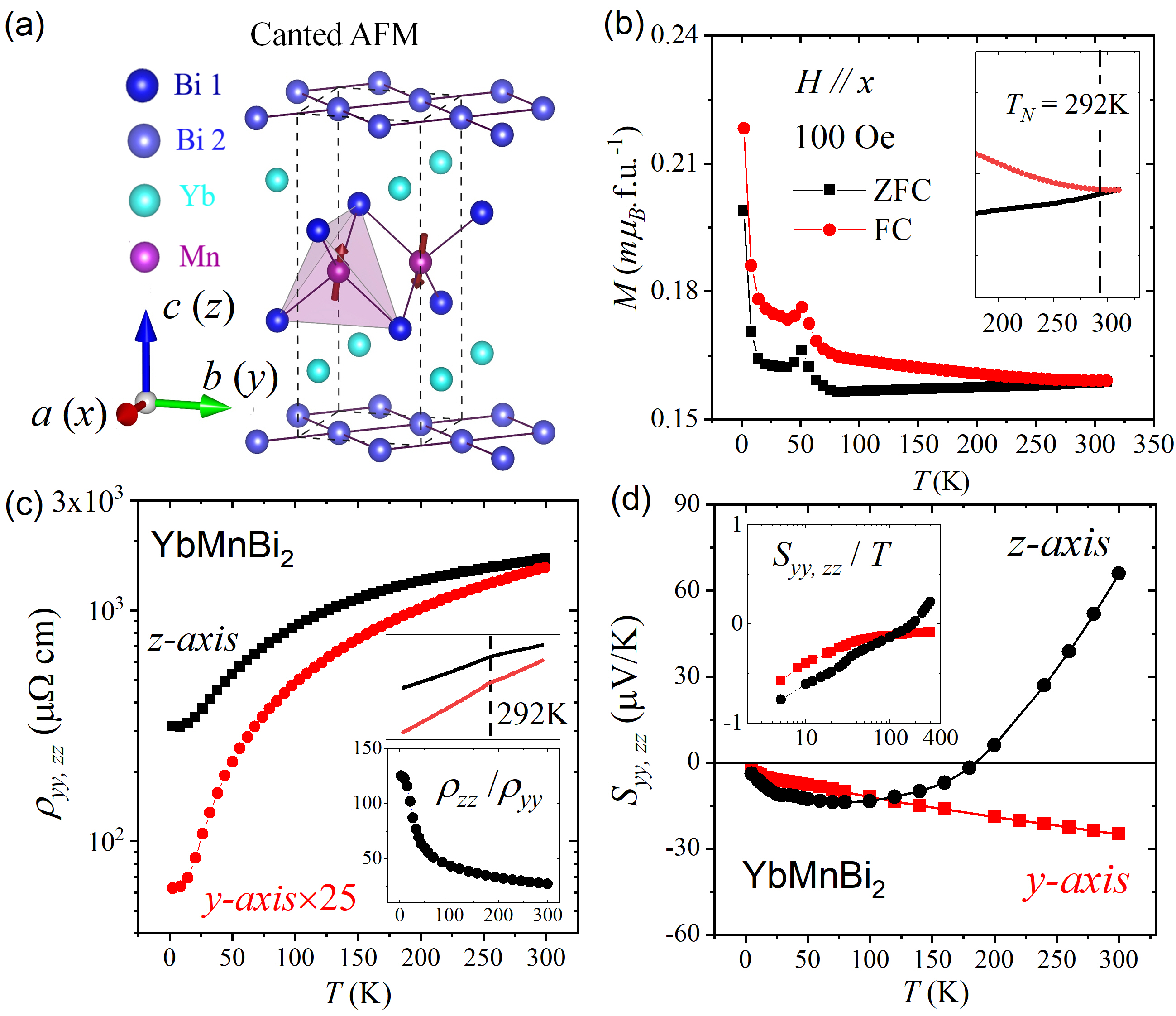}
\textcolor{black}{\caption{ \textbf{The basic properties of YbMnBi$_2$.} (a) The crystal structure of YbMnBi$_2$ with two sets of Bi: One kind of Bi is bonded to four Mn atoms while another forms a 2D intercalation network. The spin of Mn cants in the $xy$ plane.
(b) Temperature dependence of magnetization $M$ in the $xy$ plane ($H$$\|$$xy$).
(c) Temperature dependence of longitudinal resistivity $\rho_{yy}$ (red circles) and $\rho_{zz}$ (black squares).
(d) Temperature dependence of Seebeck coefficients $S_{yy}$ (red squares) and $S_{zz}$ (black circles).
 }
\label{fig:1}}
\end{figure}

\begin{figure*}[ht]
\centering
\setlength {\belowcaptionskip} {-0.4cm}
\includegraphics[scale=0.5]{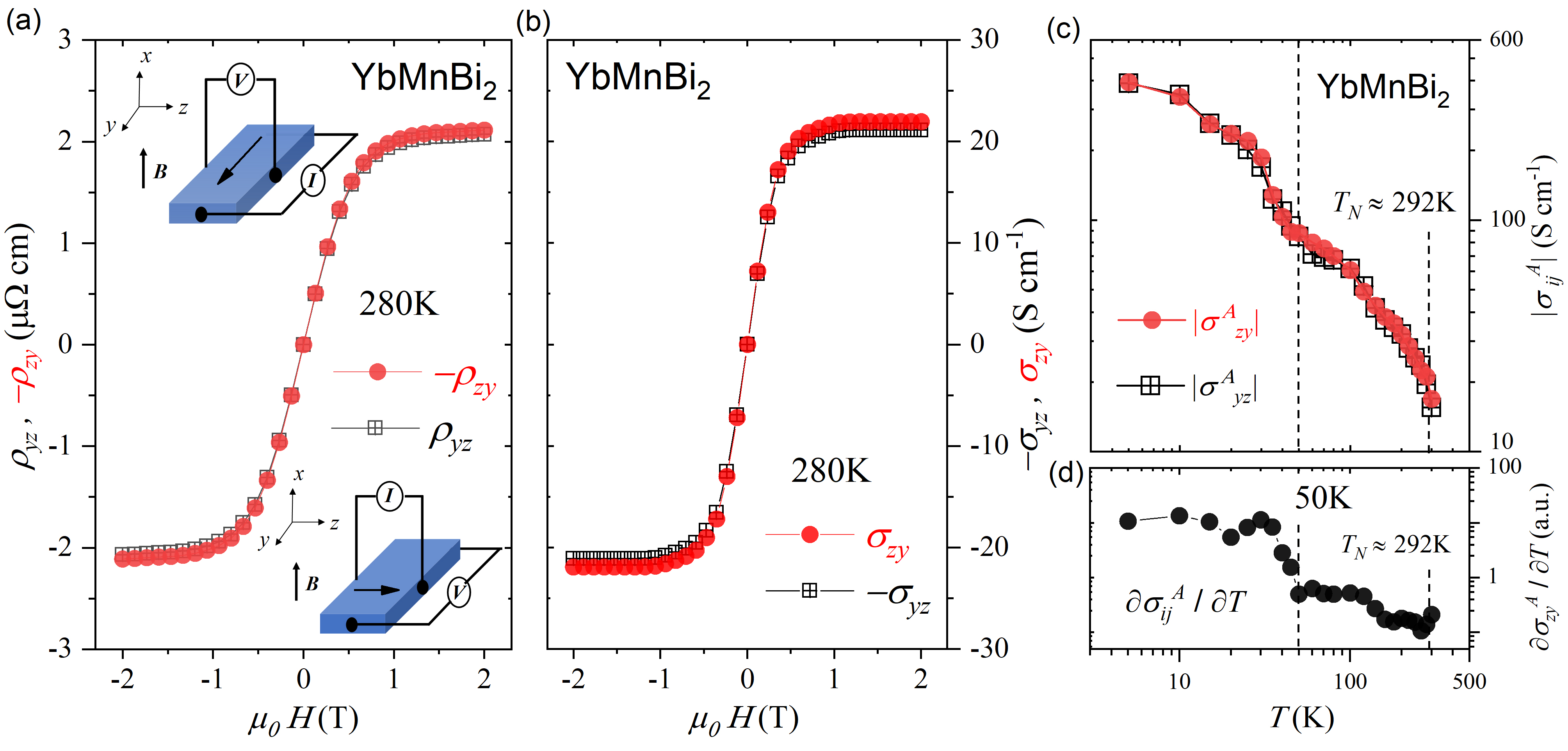}
\textcolor{black}{\caption{\textbf{Transverse electric transport.}
(a) The setup for measuring Hall signals in the $zy$ and $yz$ configurations. The Hall resistivity is identical: $\rho_{zy} = -\rho_{yz}$.
(b) The AHC $\sigma_{zy} \left(\sigma_{yz}\right)$ at 280 K.  The Hall conductivity is also identical: $\sigma_{zy} = -\sigma_{yz}$.
(c) The temperature dependence of anomalous Hall conductivity shows that Onsager's reciprocal is true in the whole temperature range. It exhibits turning an anomaly around 50 K, concomitant with an anomaly in magnetization. 
(d) The temperature derivative of $\sigma_{ij}^A / T$, which shows a kink around 50 K.}
\label{fig:2}}
\end{figure*}

YbMnBi$_2$ crystallizes in a $P4/nmm$ structure, as shown in Fig.\ref{fig:1}(a)\cite{Wang2016,pan2022giant, soh2019magnetic}. In an ordered state, the spin of Mn aligns antiferromagnetically along the $z$-axis but with a canted angle that results in a ferromagnetic component in the $xy$ plane \cite{pan2022giant, borisenko2019time,soh2019magnetic,Le2021}. This canted ferromagnetic component lifts the degeneracy of Dirac points and creates Weyl points near the Fermi level \cite{borisenko2019time, pan2022giant}. Both anomalous Hall  (AHE) and anomalous Nernst  effects were observed in this system \cite{pan2022giant}. The maximum Nernst thermopower was found to be remarkably large ($\sim 6~ \mu$V\,K$^{-1}$ at 160 K), exceeding the Nernst signal observed in other topological antiferromagnets, such as Mn$_3$Sn \cite{Li2017,Chen2022, Ikhlas2017}($\sim0.5~ \mu$V\,K$^{-1}$), in Mn$_3$Ge \cite{xu2020-2, Wuttke2019,Chen2022}($\sim1.2~\mu$V\,K$^{-1}$). 

However, Pan \textit{et al.} \cite{pan2022giant} report an intriguing breakdown of Onsager's reciprocal relations in both Hall and Nernst conductivities. According to their data (Fig. 4 in Ref. \cite{pan2022giant}),  
$\sigma_{bc}\ne -\sigma_{cb}$ and $\alpha_{bc}\ne -\alpha_{cb}$ in YbMnBi$_2$, over a wide temperature range.
This is surprising, because in contrast to, say,  the Wiedemann-Franz (WF) law, these relations are a cornerstone of irreversible thermodynamics. It has been checked experimentally \cite{xuliangcai2020} that, even when the WF law breaks down, the Bridgman relation, a consequence of Onsager reciprocity \cite{callen1948application}, holds. 

The crystal symmetry determines the number of the distinct components of the magnetoconductivity tensors \cite{akgoz-1,akgoz-2}. There are cases where components of this tensor ($\rho_{ij}$, $\sigma_{ij}$, or $\alpha_{ij}$,) can have both even (symmetric) and odd (antisymmetric) terms in a magnetic field. The most famous example of this is bismuth and its so-called ``Umkehr'' effect \cite{MICHENAUD1970455,Spathelf2022}. However, the point group  crystallographic symmetry of YbMnBi$_2$ rules this out ( see Supplementary Note 3 \cite{SM}). 

Here, we report on a careful study of YbMnBi$_2$ with the aim of quantifying the $zy$($cb$) and $yz$($bc$) components of electric and thermoelectric conductivity tensors. We find that the Hall response obeys Onsager's reciprocal relation. That is, $\rho_{zy}(H) = \rho_{yz}(-H)$ and $\sigma_{zy}(H) = \sigma_{yz}(-H)$. Onsager's reciprocal relations are also verified for the Nernst response: $\alpha_{zy}(H) = \alpha_{yz}(-H)$. On the other hand,  and as expected, in the case of Nernst thermopower: $S_{zy}(H) \ne S_{yz}(-H)$.
We also examine the temperature dependence of the anomalous transverse thermoelectric response $\alpha_{ij}^A / \sigma_{ij}^A$ ratio\cite{xuliangcai2020} and find that this ratio attains a record value of $\approx 2.9 k_B/e$ in this system.

Figure\ref{fig:1}(b) shows the temperature dependence of $M/H$ observed under $H = 100$ Oe. The N$\acute{\text{e}}$el temperature $T_\text{{N}}$ is $\sim$ 292 K \cite{soh2019magnetic,pan2022giant} revealed by a separation point between the field cooling (FC) and the zero field cooling (ZFC), which agrees with previous reports $\sim$ 290 K \cite{pan2022giant, Wang2016, borisenko2019time}. At $T\approx 50$ K, we detect, for both orientations of temperature sweep, an anomaly not detected before.  It indicates a change in the spin canting orientation below and above this temperature. 

As illustrated in Fig.\ref{fig:1}(c), resistivity is metallic and anisotropic. Both $\rho_{yy}$ and $\rho_{zz}$ exhibit a small kink near the N$\acute{\text{e}}$el temperature. The  resistivity anisotropy is $\approx 25$ at room temperature and constantly amplifies with cooling, becoming $\approx 125$ at low temperature [See the inset in Fig.\ref{fig:1}(c)]. This indicates that not only the Fermi velocity is anisotropic, but also the relative weight of different carriers and/or scattering mechanisms changes with cooling. 

The Seebeck coefficient is also anisotropic as shown in Fig.\ref{fig:1}(d). $S_{yy}$ is monotonic with temperature. In contrast, $S_{zz}$ is nonmonotonic with a peak around 70 K and a sign change above 180 K. This indicates that $S_{zz}$ has two components with different signs and different variations with temperature. Like many other anisotropic conductors, such as cuprates\cite{Silk2009} and ruthenates\cite{Daou}, the Seebeck coefficient, anisotropic at high temperatures, becomes almost isotropic at low temperatures. This is what is expected when the rough magnitude of the (normalized) Fermi energy sets the amplitude of the Seebeck coefficient despite the presence of carriers of both signs \cite{Kamran_Thermoelectricity_2004}. 
 
\begin{figure*}[ht]
\centering
\setlength {\belowcaptionskip} {-0.4cm}
\includegraphics[scale=0.5]{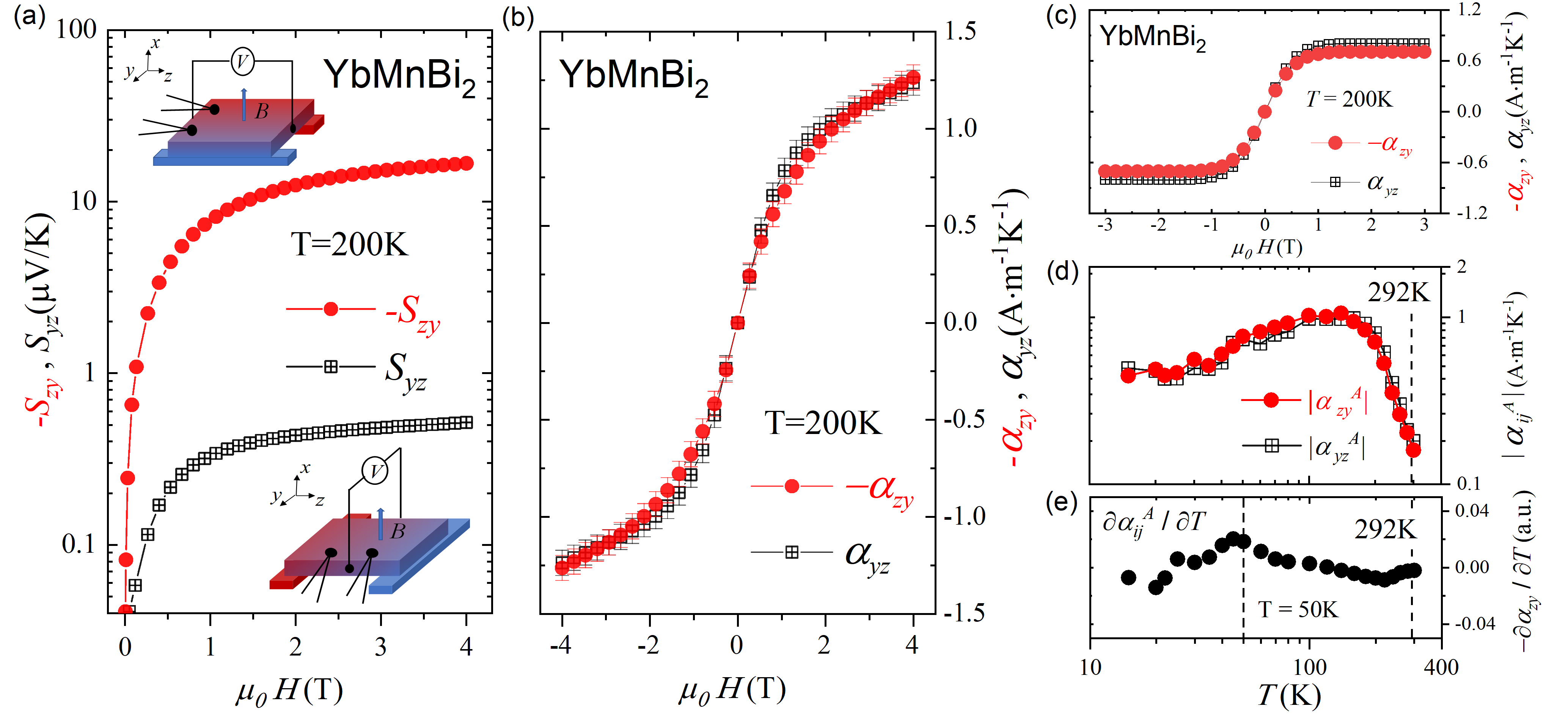}
\textcolor{black}{\caption{\textbf{Transverse thermoelectric transport.}
(a) The setup for measuring Nernst signals in the $zy$ and $yz$ configurations. The transverse thermopower shows an extreme anisotropy in both reciprocal configurations.
(b) The anomalous Nernst conductivity is calculated. Its magnitude for the two configurations is identical within experimental error shown the established Onsager's reciprocal relation.
(c) The anomalous Nernst conductivity (ANC) $\alpha_{zy} \left(\alpha_{yz}\right)$ at 200 K.
(d) The temperature dependence of ANC shows that Onsager's reciprocal is true in the whole temperature range. It also exhibits a kink around 50 K. 
(e) The temperature derivative of $\alpha_{ij}^A / T$, which shows a kink around 50 K. }
\label{fig:3}}
\end{figure*}

Given the large anisotropy between in-plane ($\rho_{yy}$) and out-of-plane ($\rho_{zz}$) resistivity, seen in Fig.\ref{fig:1}(c), one may wonder about the fate of Onsager's reciprocal relation in the Hall response. As shown in Fig.\ref{fig:2}(a), it holds. The Hall resistivity, when the magnetic field is along the $x$-axis, is identical when it is measured in the $zy (yz)$ configuration. Here $zy (yz)$ corresponds to a  current applied parallel to the $y (z)$ axis and a Hall voltage detected in the $z (y)$ direction. Thus, $\rho_{zy}(H)=\rho_{yz}(-H)$. The field sweep reveals a jump representative of the anomalous Hall effect \cite{nagaosa2010.82.1539, Li2017, sakai2018giant, xu2020-2, Nakatsuji2015, Taguchi2001}. The Hall conductivity is calculated using $\sigma_{ij} \approx \frac{-\rho_{ij}}{\rho_{ii} \rho_{jj}}$ in Fig.\ref{fig:2}(b). As shown in Fig.\ref{fig:2}(c), the two configurations yield identical values for anomalous Hall conductivity (AHC): $\sigma_{zy}^A(H) = \sigma_{yz}^A(-H)$. This equality holds for the whole temperature range [Fig.\ref{fig:2}(c)]. 

Thus, Onsager's reciprocal relation is strictly satisfied for  Hall resistivity and Hall conductivity, in the presence of a very anisotropic Fermi surface, a magnetic order, and a temperature-dependent anisotropy. The origin of the observation reported in Ref.\cite{pan2022giant} is yet to be understood. As discussed in the details in Supplemental Note 5 \cite{SM}, it might be caused by the oxidation of the samples between two sets of measurements.

Hall conductivities exhibit anomalies around 50 K [Fig.\ref{fig:2}(c)], consistent with the anomaly seen in the magnetization data shown in Fig.\ref{fig:1}(b). To further illustrate this point, we took the derivative of $\sigma_{ij}^A$, as depicted in Fig.\ref{fig:2}(d), and identified a distinct anomaly occurring at  50 K. The longitudinal transport does not show any anomaly at 50 K [Fig.\ref{fig:1}(c)]. Therefore, what happens at this temperature concerns the orientation of spins. The Fermi surface or other electronic charge-related properties are not visibly affected. 

Let us now examine the relevance of Onsager's reciprocity to the thermoelectric transport. In contrast to its electric counterpart, the anomalous Nernst effect is driven by a statistical force \cite{xiao2006berry, xiao1959-2007}, and the issue deserves to be addressed by the experiment.   In Fig.\ref{fig:3}(a), we show the results of Nernst experiment for two configurations. In the absence of charge current, when a temperature gradient $\nabla T$,  is applied, along the $y$ ($z$) axis, it gives rise to a voltage along the  $z$ ($y$) orientation. Thus, one can compare the responses for  $zy$ and $yz$ configurations. As shown in Fig.\ref{fig:3}(a), $S_{zy} \neq -S_{yz}$. At 1 T, the anisotropy is as large as 24. Since S$_{ij}$ is \textit{not} a true Onsager coefficient,  this is not a surprising result. On the other hand, $\alpha$, the thermoelectric conductivity tensor, is an Onsager coefficient which links a  force (either the temperature gradient or the electric field) to a flux (the charge density current or the heat density current). The Nernst conductivity is the off-diagonal component of this Onsager tensor: $\alpha_{ij} = S_{ij}\sigma_{ii}+S_{jj}\sigma_{ij}$. It is shown in Fig.\ref{fig:3}(b). One can see that $\alpha_{zy} = -\alpha_{yz}$ and they both show a jump during the field sweep near zero magnetic field as observed in numerous magnets \cite{pan2022giant, xu2020-2, Li2017, xuliangcai2020, Hanasaki2008,Miyasato2007,Pu2008}, shown in Fig.\ref{fig:3}(c). Its magnitude for the two configurations is identical within experimental uncertainty: $\alpha_{zy}^A(H) = \alpha_{yz}^A(-H)$. We found that this equality holds for the whole temperature range [Fig.\ref{fig:3}(d)].

Onsager's reciprocal relations lead to the following (details in Supplemental Note 6 \cite{SM}):
\begin{equation} \label{eq2}
\abovedisplayshortskip=8pt
\belowdisplayshortskip=8pt
\abovedisplayskip=8pt
\belowdisplayskip=8pt
\frac{S_{ij}}{S_{ji}} \approx -\frac{\sigma_{jj}}{\sigma_{ii}}. 
\end{equation}


\begin{figure*}[ht]
\centering
\setlength {\belowcaptionskip} {-0.4cm}
\includegraphics[scale=0.6]{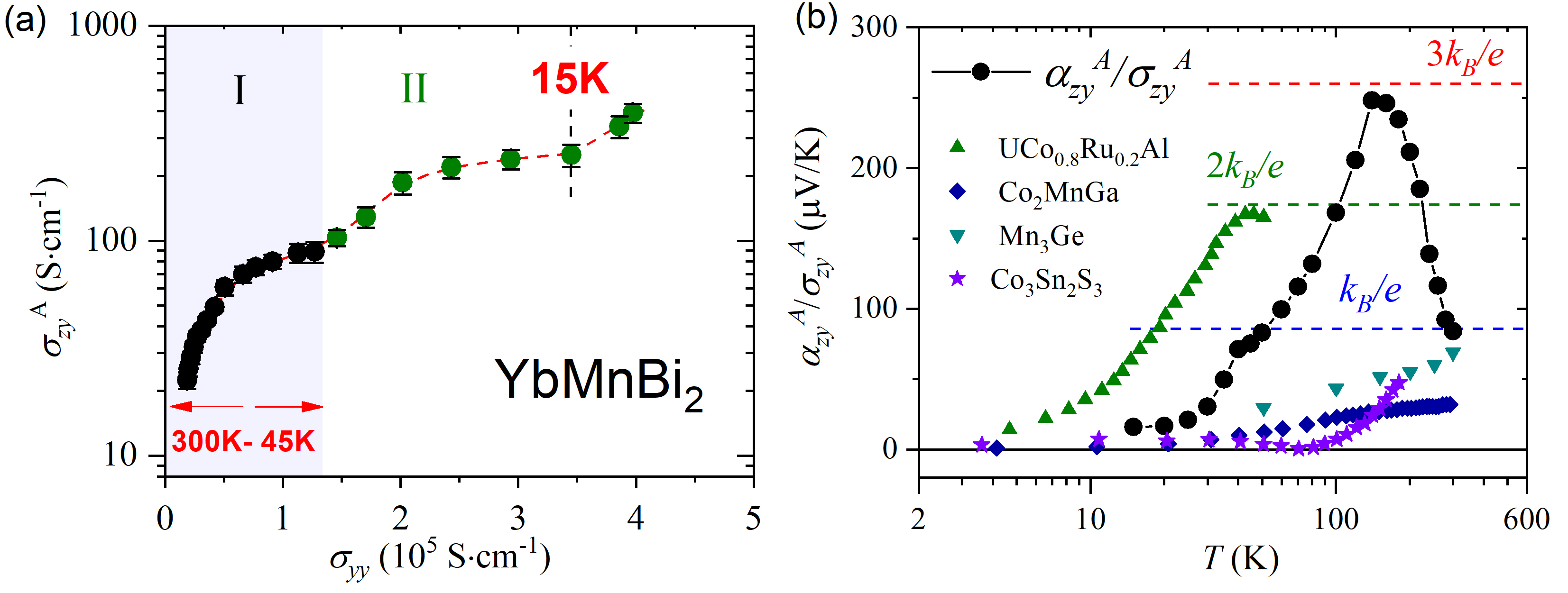}
\textcolor{black}{\caption{\textbf{Scaling relation of the anomalous Hall conductivity.} The longitudinal conductivity (a) $\sigma_{yy}$ for YbMnBi$_2$ lines within the good-metal regime, and there is a crossover behavior to the skew scattering region in the lower temperature, suggesting
the skew scattering will dominate the transverse transport. The flat AHC [$\sigma_H^A$ = $(\sigma)^0 =$ const] is concomitant with the maximum of magnetization at 50 K in region I. A second observed intrinsic AHC can be attributed to the magnetization turning point at 15 K in region II. An extrinsic mechanism may play a role by below 15 K. (b) The $\alpha_{zy}^A/\sigma_{zy}^A$ ratio at different temperatures. Similar to other topological magnets, it approaches 86 $\mu$V/K at room temperature, but there is a peak as large as 250 $\mu$V/K at 150 K.}
\label{fig:4}}
\end{figure*}
As in the case of the anomalous Hall effect, there is a kink in $S^A_{ij}$ and $\alpha^A_{ij}$ [Fig.\ref{fig:3}(d) and Supplemental Note 6 \cite{SM}], near the temperature at which there is an anomaly in magnetization. A more visible anomaly can be seen in the temperature derivative of $\alpha^A_{ij}$ as presented in Fig.\ref{fig:3}(e). The Seebeck coefficient does not show any anomaly at 50 K [Fig.\ref{fig:1}(d)]. This further suggests that what occurs at 50 K is a spin-related and not a charge-related phenomenon. 

Since Onsager's reciprocity is found to hold in the whole temperature range, let us demonstrate that this implies its validity irrespective of the origin of the anomalous transverse coefficients.  Figure\ref{fig:4}(a) shows how the amplitude of the anomalous Hall conductivity for the $yz$ configuration varies as a function of longitudinal conductivity as the sample is cooled.  Two regions, I and II can be distinguished in this plot. In region I, at high temperature above the anomaly in magnetization, $\sigma^A_{zy}$ increases with increasing $\sigma_{yy}$ and then saturates to a plateau near 50 K. This flat AHC [$\sigma_H^A$ = $(\sigma)^0 =$ const] is concomitant with the maximum of magnetization at 50 K and may be attributed to an intrinsic AHC \cite{sakai2018giant, Guin2018AnomalousNE, xuliangcai2020, xu2022topological, Onoda2006, nagaosa2010.82.1539}. With decreasing temperature and entrance to region II, magnetization increases to a turning point at 15 K, and the AHC also saturates near the same temperature, representing a second intrinsic AHC. A transformation of the intrinsic Berry curvature at 50 K would explain the observation of two saturated AHC amplitudes around 50 and 15 K. Below 15 K, $\sigma^A_{zy}$ increases again as a function of $\sigma_{yy}$, indicating a role played by extrinsic (skew scattering or side jump) mechanism of AHE \cite{nagaosa2010.82.1539,Onoda2006,Tian2009}. The change in the in-plane ferromagnetic component at 50 K affects the intrinsic Berry curvature. Density functional theory (DFT) \cite{P2022} finds that the number of the Weyl node pairs will increase with increasing canted angle.

Given the presence of both intrinsic and extrinsic components in AHE and the equality between off-diagonal components of AHC and ANC in the whole temperature range, we can conclude that Onsager's reciprocity is relevant to anomalous transverse response regardless of the details of their microscopic origin.

Let us now turn our attention to the relative amplitude of $\sigma_{ij}^A$ and $\alpha_{ij}^A$. Xu \textit{et al}. \cite{xuliangcai2020} found that the ratio of the $\alpha_{ij}^A$/$\sigma_{ij}^A$ approaches the ratio of natural units $k_B/e$ = 86 $\mu$V\,K$^{-1}$ at room temperature in many topological magnets. The temperature dependence of the $\alpha_{ij}^A$/$\sigma_{ij}^A$ ratio is a monotonic function of the temperature and approaches $k_B/e$ near the ordering temperature \cite{ Guin2018AnomalousNE, xu2022topological,ding2019, zhang2022exchange}. In contrast, as shown in Fig.\ref{fig:4}(b), we find that, in YbMnBi$_2$, the temperature dependence of this ratio is nonmonotonic and attains a value as large as 2.9 $k_B/e$ at 160 K.

In the intrinsic picture, this ratio depends on the way the Berry curvature affects the flow of entropy and the flow of charge \cite{ding2019, xuliangcai2020}. Roughly, the ratio is set by $\frac{k_B}{e}  \frac{\langle\lambda_F^2\rangle}{\langle\Lambda^2\rangle} $ \cite{xuliangcai2020}, where $\langle\lambda_F^2\rangle$ and $\langle\Lambda^2\rangle$ are the square of the Fermi and the thermal wavelengths averaged over the whole Fermi surface \cite{xuliangcai2020}, respectively. As the system is warmed up, the  $\alpha_{ij}^A$/$\sigma_{ij}^A$ ratio monotonically increases and tends toward  $\sim \frac{k_B}{e}$ when $\lambda_{F}$ and $\Lambda$ become comparable.  A number of theoretical studies have examined the amplitude of this ratio \cite{Zhiqiang2022,Lu2023}. Qiang \textit{et al.} \cite{Lu2023}, by performing a Sommerfeld expansion, found that this ratio becomes $\frac{\alpha_{ij}^{int}}{\sigma_{ij}^{int}} = \left(\frac{\mu}{e} + \frac{\pi^2}{3}\frac{k_B^2T^2}{e\mu}\right)^{-1}L_0T$. This expression implies an upper boundary of $\sim 0.77 \frac{k_B}{e}$ \cite{Lu2023,ashcroft2011solid}(see Supplemental Note 7 \cite{SM}). Therefore, our observation calls for an approach beyond a single-band degenerate Fermi system subject to Sommerfeld expansion.


The Fermi surface of YbMnBi$_2$ is known to consist of multiple electron and hole pockets. According to the most recent set of DFT calculations \cite{pan2022giant}, the largest pockets are electron-like and have a  Fermi energy in the range of  $\sim$80 meV \cite{pan2022giant, borisenko2019time, Wang2016}. This is confirmed by our quantum oscillation (SdH) results (see Supplemental Note 8 \cite{SM} ) and consistent with the slope of our isotropic thermopower at low temperatures shown in Fig.\ref{fig:1}(d).  $S_{ii}/T \rightarrow 0.6 \mu $V/K$^2$ implies a Fermi energy of $\sim 50$ meV. Since this is the largest energy scale of the system, the Fermi energy of other pockets (and, in particular, the smaller hole pockets) should be significantly smaller than this ,and, at 150 K,  at least one set of hole pockets are nondegenerate.
 
There is no available theory of intrinsic anomalous Hall effect in the presence of nondegenerate electrons. Nevertheless, let us note that a nondegenerate electron has more entropy than a  degenerate electron \cite{Collignon2020}, but the same electric charge. Therefore, it is plausible that the presence of nondegenerate carriers allows a larger $\alpha_{ij}^A$/$\sigma_{ij}^A$ ratio. Let us also not forget that the $k_B/e$ boundary may not hold in the presence of multiple bands or when AHE is partially extrinsic. 

In summary, we measured electric and thermoelectric transport properties in YbMnBi$_2$ and found that Onsager's reciprocal relation is robust. Reciprocity holds in the whole temperature range irrespective of the intrinsic or extrinsic origin of the anomalous transverse response. The  $\alpha_{ij}^A$/$\sigma_{ij}^A$ ratio is exceptionally large, possibly as a result of the presence of nondegenerate electrons.     

This work was supported by The National Key Research and Development Program of China (Grant No. 2022YFA1403500), the National Science Foundation of China (Grants No. 12004123, No. 51861135104, and 11574097 ), and the Fundamental Research Funds for the Central Universities (Grant No. 2019kfyXMBZ071). K. B. was supported by the Agence Nationale de la Recherche (ANR-19-CE30-0014-04). X. L. acknowledges the China National Postdoctoral Program for Innovative Talents (Grant No. BX20200143) and the China Postdoctoral Science Foundation (Grant No. 2020M682386).
* \verb|zengwei.zhu@hust.edu.cn|\\
$\dagger$ \verb|kamran.behnia@espci.fr|\\

%

\providecommand{\noopsort}[1]{}\providecommand{\singleletter}[1]{#1}%

\clearpage
\onecolumngrid
\begin{center}
	{\large Supplemental Material for\\[2mm] \bf Onsager Reciprocal Relation between Anomalous Transverse Coefficients of an Anisotropic Antiferromagnet}\\[3mm]
	Xiaodong Guo$^{1}$, Xiaokang Li$^{1}$, Zengwei Zhu$^{1,*}$ and Kamran Behnia$^{2,\dagger}$\\[2mm]
	{\small {\it $^1$Wuhan National High Magnetic Field Center and School of Physics, Huazhong University of Science and Technology,  Wuhan,  430074, China\\
			$^2$Laboratoire de Physique et Etude des Mat\'{e}riaux (CNRS/UPMC),Ecole Sup\'{e}rieure de Physique et de Chimie Industrielles, 10 Rue Vauquelin, 75005 Paris, France}\\[0mm]}
\end{center}
\setcounter{page}{1}
\vspace*{5mm}


\maketitle

\twocolumngrid
\renewcommand{\thefigure}{S\arabic{figure}}
\renewcommand{\thetable}{S\arabic{table}}
\def\theequation{S\arabic{equation}}

\makeatletter
\def\@hangfrom@section#1#2#3{\@hangfrom{#1#2}#3}
\def\@hangfroms@section#1#2{#1#2}
\makeatother

\renewcommand{\thesection}{S\arabic{section}}
\renewcommand{\thetable}{S\arabic{table}}
\renewcommand{\thefigure}{S\arabic{figure}}
\renewcommand{\theequation}{S\arabic{equation}}

\setcounter{section}{0}
\setcounter{figure}{0}
\setcounter{table}{0}
\setcounter{equation}{0}
\maketitle

\section*{SUPPLEMENTARY NOTES}

\subsection*{ Supplementary Note 1. Single-crystal preparation and methods}
The YbMnBi$_2$ single crystals were synthesized using a self-flux method with the stoichiometric mixture of Yb, Mn, and Bi elements with an elemental ratio of Yb:Mn:Bi of 1:1:6. The starting materials were put into a small alumina crucible and sealed in a quartz tube in Argon gas atmosphere. The tube was then heated to 1050 °C for 2 days, followed by a subsequently cooling down to 400 °C at a rate of 2 °C h$^{-1}$. The plate-like single crystals as large as a few millimeters can be obtained.

\begin{figure}[ht]
	\setlength {\abovecaptionskip} {-0.2cm}
	\setlength {\belowcaptionskip} {-0.4cm}
	\includegraphics[width=9cm]{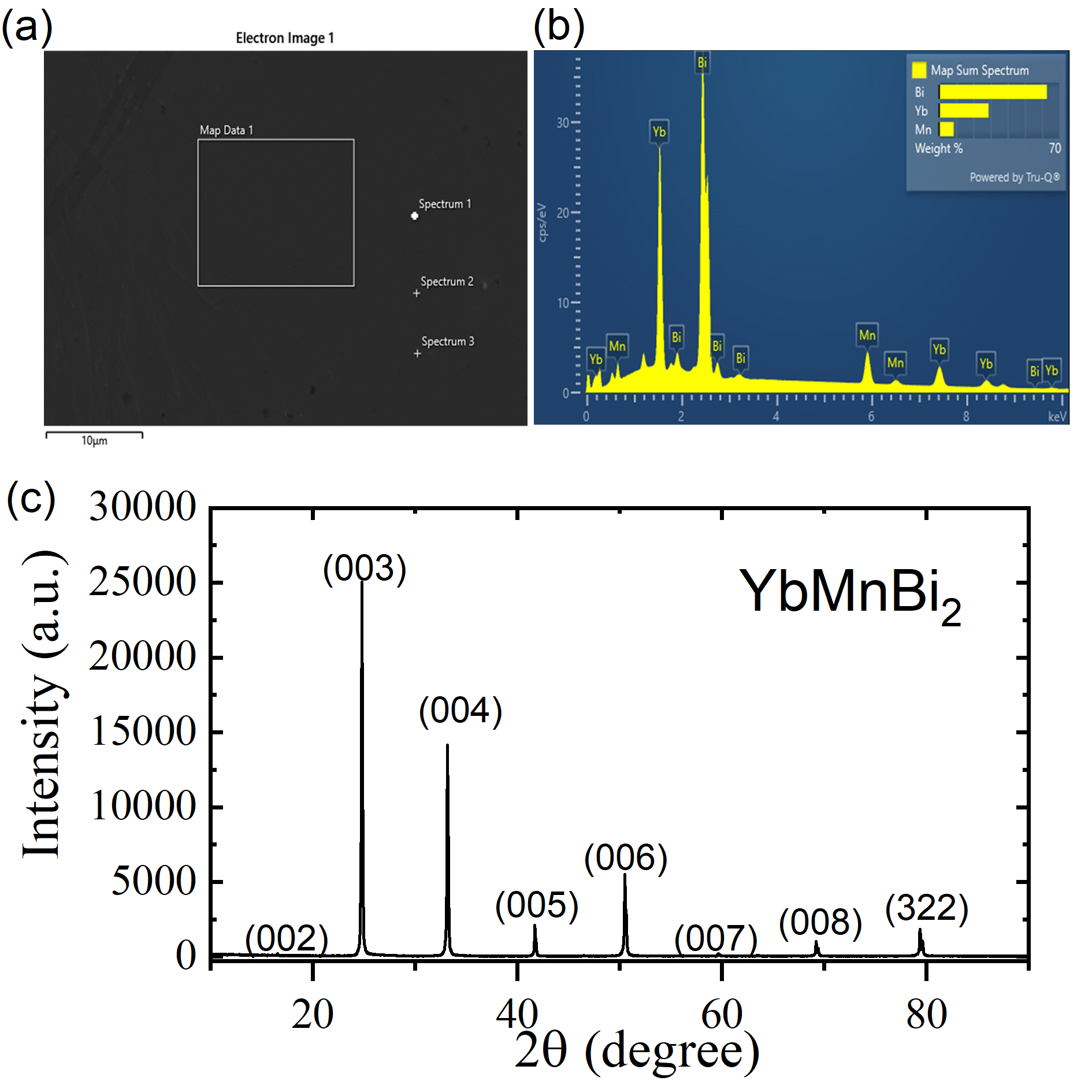}
	\textcolor{black}{\caption{ (a) Scanning electron microscope (SEM) images for the samples discussed in the main text. (b) Energy dispersive spectroscopy (EDS) spectrum of map data of the samples. (c) the XRD diffraction spectrum of YbMnBi$_2$ sample in our work.
		}
		\label{fig:S1}}
\end{figure}

The samples discussed in the main text were also examined by energy dispersive spectroscopy (EDS) to determine their compositions. The results are listed in Fig. \ref{fig:S1} and table \ref{table:S1}. Also, the single crystal XRD is carried out to analyze the structure. The diffraction spectrum has been shown in Fig.\ref{fig:S1}. The lattice parameter can be obtained and summarized in table. 

The experimental methods were the same as the previous reports\cite{li2017sm,xuliangcai2020sm,xu2020-2sm,ding2019sm}.

\begin{table}[ht]   
	\begin{center} 
		\caption{Elemental ratios in YbMnBi$_2$ samples from EDS, showing a consistent chemical composition result.}
		\label{table:S1} 
		\begin{tabular}{|m{2cm}<{\centering}|m{2.2cm}<{\centering}|m{2.2cm}<{\centering}|m{2.2cm}<{\centering}|}   
			\hline   \textbf{Spectrum Label} & \textbf{Yb (Atomic \%)} & \textbf{Mn (Atomic \%)} & \textbf{Bi (Atomic \%)}  \\   
			\hline   Map Data 1 & 25.36 & 25.16 & 49.48 \\ 
			\hline   Spectrum 1 & 25.67 & 24.94 & 49.39 \\  
			\hline   Spectrum 2 & 26.06 & 25.39 & 48.55 \\  
			\hline   Spectrum 3 & 25.36 & 25.15 & 49.49 \\ 
			\hline   
		\end{tabular}   
	\end{center}
\end{table}

\begin{table}[ht]   
	\begin{center} 
		\caption{Lattice parameter of YbMnBi$_2$ obtained by single crystal XRD diffraction.}
		\label{table:S2} 
		\begin{tabular}{|m{2.6cm}<{\centering}|m{1.6cm}<{\centering}|m{1.6cm}<{\centering}|m{1.6cm}<{\centering}|}   
			\hline   \textbf{Sample label} & \textbf{$a$ ($\AA$)} & \textbf{$b$($\AA$)} & \textbf{$c$($\AA$)}  \\   
			\hline   Sample 1 & 4.472 & 4.502 & 10.803 \\ 
			\hline   Sample 2 & 4.478 & 4.51 & 10.717 \\  
			\hline   Sample 2 & 4.474 & 4.501& 10.758 \\  
			\hline   Ref.1 & 4.48 & 4.48 & 10.8 \\ 
			\hline   
		\end{tabular}   
	\end{center}
\end{table}

\subsection*{ Supplementary Note 2. Sign convention of transverse transport measurements}

\begin{figure}[ht]
	\setlength {\abovecaptionskip} {-1cm}
	\setlength {\belowcaptionskip} {-0.8cm}
	\includegraphics[width=9cm]{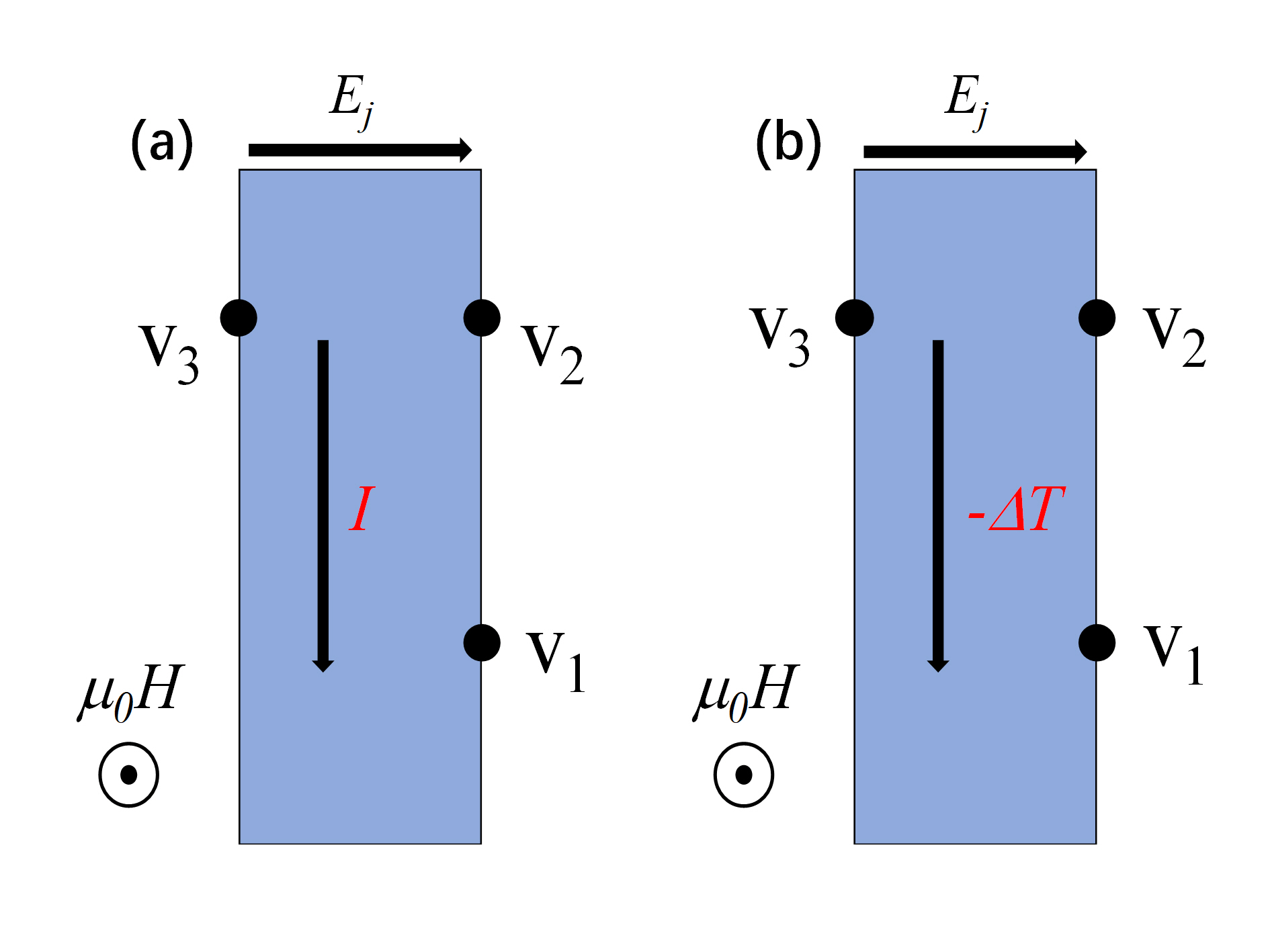}
	\textcolor{black}{\caption{ Sign definition for Hall (a) and Nernst signals (b) in our experimental setup.}
		\label{fig:S2}}
\end{figure}

Here we elaborate on the sign convention of transverse transport measurements. For this purpose, Supplementary Fig.\ref{fig:S2} shows the schematics of experimental device. 

i). $\rho_{ii}$, $\sigma_{ii}$ and $S_{ii}$: for $\rho_{ii}$, $\sigma_{ii}$, the sign is always positive. For $S_{ii}$, the sign is defined as  
\begin{equation}
	\abovedisplayshortskip=8pt
	\belowdisplayshortskip=8pt
	\abovedisplayskip=8pt
	\belowdisplayskip=8pt
	S_{ii} = - \frac{\Delta V_{ii}}{\Delta T_{ii}} = - \frac{V_2-V_1}{\Delta T_{ii}}.
\end{equation}

ii). Hall effect $\rho_{ij}$: A the schematic figure is shown in Fig. \ref{fig:S2}(a). We define $\rho_{ji} = \frac{V_{ij}}{I} = \frac{V_3 - V_2}{I}$. where $I$ is the magnitude of the electric current passing through the sample.

iii). Nernst effect $S_{ij}$: As shown in Fig. \ref{fig:S2}(b). We define Nernst coefficient’s sign is $S_{ij} = \frac{(V_3 - V_2)/w}{\Delta T_{ii}/l}$.  

\subsection*{ Supplementary Note 3. The form of the transverse transport coefficients}

\begin{figure}[ht]
	\setlength {\abovecaptionskip} {-0.4cm}
	\setlength {\belowcaptionskip} {-0.80cm}
	\includegraphics[width=9cm]{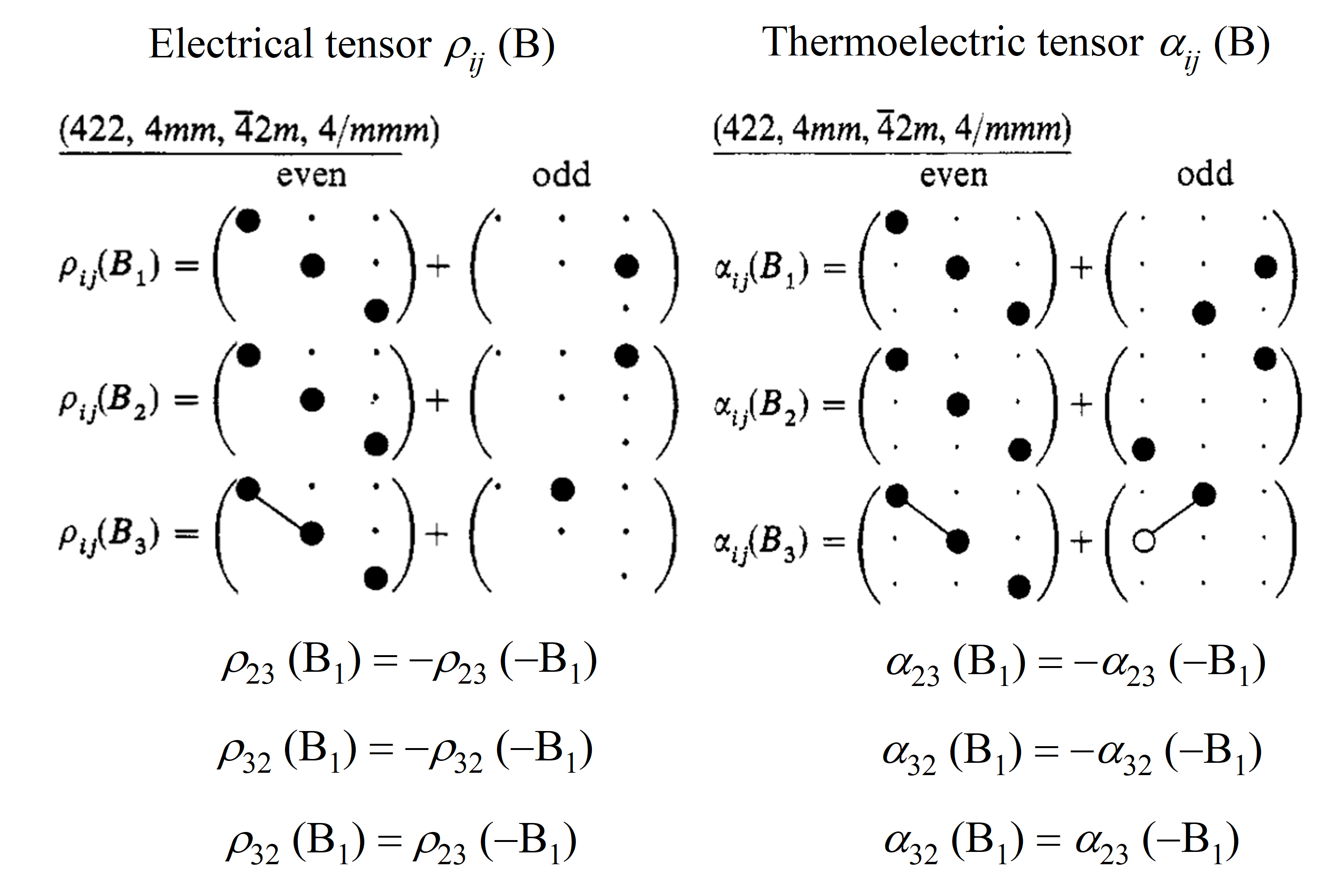}
	\textcolor{black}{\caption{ The form of transverse electrical tensor (a) and thermoelectric tensor (b) in 4/mmm point group\cite{akgoz-1sm,akgoz-2sm}. Where the big solid dots represent the non-zero component and the dots represent the zero component. All even parts and odd parts are symmetric and antisymmetric about their leading diagonals, respectively. }
		\label{fig:S3}}
\end{figure}

Akgoz \textit{et al.} has demonstrated that the form of the magnetoresistivity tensor $\rho_{ij} (B)$ and the magnetothermoelectric power tensor $\alpha_{ij} (B)$ depends on symmetry operators of the 32 crystallographic point groups\cite{akgoz-1sm,akgoz-2sm}. Each component of transport coefficients can have even and odd terms in the magnetic induction $B$. The crystal symmetry sets the number of independent odd and even components of the magnetic transport tensor. In the case of YbMnB$_2$, whose point group symmetry is 4/$mmm$. As shown in Fig.\ref{fig:S3}, the electrical and thermoelectric components have only one odd component when $B//x$. The odd components are antisymmetric in the $zy$ and $yz$ configurations. Therefore, $\rho_{zy}(B)$ = $\rho_{yz}(-B)$ and $\alpha_{zy}(B)$ = $\alpha_{yz}(-B)$.

\subsection*{ Supplementary Note 4. Magnetization measurements}

The temperature dependence of the magnetization curves is shown in Fig.\ref{fig:S4} (a) and (b) for the field parallel to the $x$ and the $z$ axes. There is a strong anisotropy in the two configurations. As shown in Fig.\ref{fig:S4} (c) and (d), magnetization shows a field dependence along the $xy$ plane. 

\subsection*{ Supplementary Note 5. Anomalous Hall effect}

The Hall resistivity was measured from 300 K to 5 K, as shown in Fig.\ref{fig:S5}. The Hall conductivity is calculated using $\sigma_{ij}= \frac{-\rho_{ij}}{\rho_{ii}\rho_{jj} + \rho_{ij}\rho_{ji} }$, and is shown in Fig.\ref{fig:S6}. At each temperature, the Hall conductivity can be empirically expressed as
\begin{equation}
	\abovedisplayshortskip=8pt
	\belowdisplayshortskip=8pt
	\abovedisplayskip=8pt
	\belowdisplayskip=8pt
	\sigma_{zy} = \sigma_{zy}^A + \sigma_{zy}^T + R_0B  = \sigma_HM + \sigma_{zy}^T + \sigma_0B.
\end{equation}
Here, $R_0$ and $R_H$ correspond to the normal and anomalous Hall coefficients, respectively. $\rho_{zy}^T$ is the topological Hall effect (THE). For example, at 120 K, the normal Hall effect is linearly fitted and subtracted in Fig.\ref{fig:S7} (f). Also, the anomalous and topological Hall signals are shown in Fig.\ref{fig:S7} (d) and (e), in which the Hall signals are identical in both reciprocal configurations and remain true in the whole temperature range as shown in Fig.\ref{fig:S7} (g), (h), and (i), respectively.

It is worth noting that YbMnBi$_2$ samples are extremely susceptible to oxidation. In Fig.\ref{fig:S8}, the Hall resistivity is different at identical temperatures for the same sample if it is stored in a poor vacuum environment for four days.

\subsection*{ Supplementary Note 6. Anomalous Nernst effect }

The experiment results of the anomalous Nernst effect (ANE) are shown in Fig.\ref{fig:S9} and Fig.\ref{fig:S11}. In the $zy$ configurations, the anomalous Nernst (ANE) is obvious from the 300 K to 10 K; however, the $S_{yz}$ suddenly appears a signal change at the 25 K. Similar to the Hall effect, the Nernst conductivities are contributed with anomalous Nernst conductivity (ANC), $\alpha_{ij}^A$, topological Nernst conductivity (TNC), $\alpha_{ij}^T$, and normal Nernst conductivity (NNC), $\alpha_0$, which can be expressed as 
\begin{equation}
	\abovedisplayshortskip=8pt
	\belowdisplayshortskip=8pt
	\abovedisplayskip=8pt
	\belowdisplayskip=8pt
	\alpha_{ij} = \alpha_{ij}^A + \alpha_{ij}^T + \alpha_0  = \alpha_HM + \alpha_{ij}^T + \alpha_0.
\end{equation}
The value of ANC is summarized in Fig.\ref{fig:S11} (f), which shows $\alpha_{zy}^A = -\alpha_{yz}^A$. As discussed in the main text, the transverse entropy flux derived by a statistical force is linked to the kinetic coefficient known as Peltier's coefficient $\alpha_{ij}$ rather than the thermopower $S_{ij}$. The topological and normal Nernst conductivity can also establish Onsager's reciprocal relation in the whole temperature range shown in Fig.\ref{fig:S11} (g) and (h).

The Nernst conductivity is the off-diagonal component of this Onsager tensor: $\alpha_{ij} = S_{ij}\sigma_{ii}+S_{jj}\sigma_{ij}$. By utilizing Onsager's reciprocal relation, we can obtain $\alpha_{zy} = -\alpha_{yz}$ in the same field. We can obtain that 
\begin{equation}\label{eqs3}
	\abovedisplayshortskip=8pt
	\belowdisplayshortskip=8pt
	\abovedisplayskip=8pt
	\belowdisplayskip=8pt
	\sigma_{yy}S_{yz}+\sigma_{yz}S_{zz}= -\sigma_{zz}S_{zy}-\sigma_{zy}S_{yy}.
\end{equation}
By utilizing Onsager's reciprocal relation again for Hall conductivity, $\sigma_{zy} = - \sigma_{yz}$, the equation(\ref{eqs3}) can rewritten as
\begin{equation}\label{eqs4}
	\abovedisplayshortskip=8pt
	\belowdisplayshortskip=8pt
	\abovedisplayskip=8pt
	\belowdisplayskip=8pt
	S_{yz}= - \frac{\sigma_{zz}}{\sigma_{yy}} S_{zy} - \frac{\sigma_{zy}}{\sigma_{yy}} (S_{yy}-S_{zz}).
\end{equation}
Here we ignore the small amount $\frac{\sigma_{zy}}{\sigma_{yy}} (S_{yy}-S_{zz})$ to obtain the following conclusion
\begin{equation} \label{eqs5}
	\abovedisplayshortskip=8pt
	\belowdisplayshortskip=8pt
	\abovedisplayskip=8pt
	\belowdisplayskip=8pt
	\frac{S_{ij}}{S_{ji}} \approx -\frac{\sigma_{jj}}{\sigma_{ii}}, 
\end{equation}
which means that the value of Nernst thermopower will be apportioned by the mobility $\mu_{ii}$, giving rise to an enhanced thermopower in one of the configurations for an anisotropic system. Note that thermopower becomes isotropic at low temperatures which results in the equation(\ref{eqs5}) more accurate.

\subsection*{ Supplementary Note 7. The $\alpha_{ij}^A / \sigma_{ij}^A$ ratio }

Qiang \textit{et al.} have recently proposed this expression for the ratio of the anomalous Nernst to anomalous Hall conductivities in the intrinsic picture\cite{Lu2023sm}: 

\begin{equation} \label{eqs6}
	\abovedisplayshortskip=8pt
	\belowdisplayshortskip=8pt
	\abovedisplayskip=8pt
	\belowdisplayskip=8pt
	\frac{\alpha_{ij}^{int}}{\sigma_{ij}^{int}} = \left(\frac{\mu}{e} + \frac{\pi^2}{3}\frac{k_B^2T^2}{e\mu}\right)^{-1}L_0T.
\end{equation}

This expression was obtained using a Sommerfeld expansion, which is valid only deep in the degenerate regime (when $\mu \ll k_BT$) \cite{ashcroft2011solidsm}. Let us write this expression in an explicit way:
\begin{equation} \nonumber
	\abovedisplayshortskip=8pt
	\belowdisplayshortskip=8pt
	\abovedisplayskip=8pt
	\belowdisplayskip=8pt
	\frac{\alpha_{ij}^{int}}{\sigma_{ij}^{int}} = \left(\frac{\mu}{e} + \frac{\pi^2}{3}\frac{k_B^2T^2}{e\mu}\right)^{-1}\frac{\pi^2}{3}\frac{k_B^2}{e^2}T 
\end{equation}

\begin{equation} \label{eqs7}
	\abovedisplayshortskip=8pt
	\belowdisplayshortskip=8pt
	\abovedisplayskip=8pt
	\belowdisplayskip=8pt
	=\frac{k_B}{e}\left(\frac{\pi^2}{3}\frac{\mu}{k_BT} + \frac{k_BT}{\mu}\right)^{-1}
\end{equation}

If $\frac{k_BT}{\mu} = 1$, then  $\frac{\alpha_{ij}^{int}}{\sigma_{ij}^{int}} =0.77 k_B/e$. However, Sommerfeld expansion is only allowed when $\frac{k_BT}{\mu} \ll 1$.
This indicates that to explain our observation one should invoke multiple bands and possibly the presence of non-degenerate carriers.

\subsection*{ Supplementary Note 8. longitudinal magnetoresistance and quantum oscillations (SdH)}

Fig.\ref{fig:S12} displays the magnetoresistance (MR) in $y$ ($I$$//$$y$ and V$//$$y$) and $z$ ($I$$//$$z$ and $V$$//$$z$) directions. The results indicate anisotropic behaviors, with the MR for $\rho_{yy}$ showing negative magnetoresistivity with increasing temperature, while $\rho_{zz}$ consistently demonstrates positive magnetoresistivity.  

Quantum oscillations in magnetoresistivity (SdH) can be observed when the magnetic field is along the $z$ axis (Fig.\ref{fig:S13}(a) and (b)). From the fast Fourier transform (FFT) background-subtracted oscillating component $\Delta\rho_{xx}$ features frequency $F_{\alpha}$ = 125 T and $F_{\beta}$ = 165 T at 3 K (Fig.\ref{fig:S13}(c)). From the Onsager relation, $F = (\Phi_{0}/2\pi^2)A_F$,  where $\Phi_{0}$ is the flux
quantum and $A_F$ is the orthogonal cross-sectional area of the Fermi surface. We estimate $A_F$ to be 1.95 nm$^{-2}$ and 1.48 nm$^{-2}$ corresponding to the measured frequencies of 165 T and 125 T, respectively. The Fermi energies of the two pockets is 102 meV and 62 meV.

In order to better reveal quantum oscillation in our sample, we extended the measurements to pulsed  magnetic field. As shown in the inset of Fig.\ref{fig:S13}(a), oscillations are observed when the field was swept up to 55 T. According to the Fast Fourier Transform (FFT) of the data, the oscillation frequency is $\sim$160 T shown in insert of Fig.\ref{fig:S13}(c), consistent with what was obtained in the static field of (165 T).

\subsection*{ Supplementary Note 9. Summary of the difference for the anomalous Hall effect and magnetization in YbMnBi$_2$} 

In order to look more carefully at the two data sets (between our work and Ref.\cite{pan2022giantsm}), we compared the anomalous Hall data as shown in Fig.\ref{fig:S15}. One can see that for one configuration, the two data sets match, but not for the other one. This suggests that the most plausible source of discrepancy is measurement or oxidation between two sets of measurement. We also compared other transverse coefficients at 200 K. This is summarized in Tab.\ref{table:S3}. Furthermore, the field dependence of magnetization at 10K is also summarized in Fig.\textcolor{blue}{S16}. The magnetization is slightly different ($\sim$5, $\sim$3.6, and $\sim$4 $m\mu_B$/f.u.) in the three sets of data from Ref.\cite{soh2019magneticsm, pan2022giantsm}and our work as a field of 1 T is applied along the $ab$-plane.

\begin{table}[h] 
	\setlength {\abovecaptionskip} {0.1cm}
	\setlength {\belowcaptionskip} {-0.1 cm}
	\begin{center} 
		\caption{The difference of transverse transport 
			coefficients between Ref.1 and our work at 200K.}
		\label{table:S3} 
		\begin{tabular}{|m{4cm}<{\centering}|m{2.5cm}<{\centering}|m{2.5cm}<{\centering}|}   
			\hline   \textbf{Transverse coefficient} & \textbf{Ref.1} & \textbf{In this work}  \\   
			\hline   $\rho_{zy}^A/\rho_{yz}^A$($\mu\Omega$cm) & $\sim$1.66 $\sim$0.14  & $\sim$1.59 / $\sim$1.51 \\ 
			\hline   $\sigma_{zy}^A/\sigma_{yz}^A$ ($Scm^{-1}$) & $\sim$1.74 / $\sim$19.98 & $ \sim$32.14/ $\sim$32.23  \\  
			\hline   $S_{zy}^A/S_{yz}^A$($\mu V/K$) & $\sim$6.0 / $\sim$3.0 &$\sim$10.74 / $\sim$0.35  \\  
			\hline   $\alpha_{zy}^A/\alpha_{yz}^A$($Am^{-1}K^{-1}$)& $\sim$0.29 / $\sim$3.47 & $\sim$0.71 / $\sim$0.78  \\ 
			\hline   
		\end{tabular}   
	\end{center}
\end{table}

\begin{figure*}[ht]
	\setlength {\abovecaptionskip} {0.2cm}
	\setlength {\belowcaptionskip} {-0.2cm}
	\includegraphics[width=13cm]{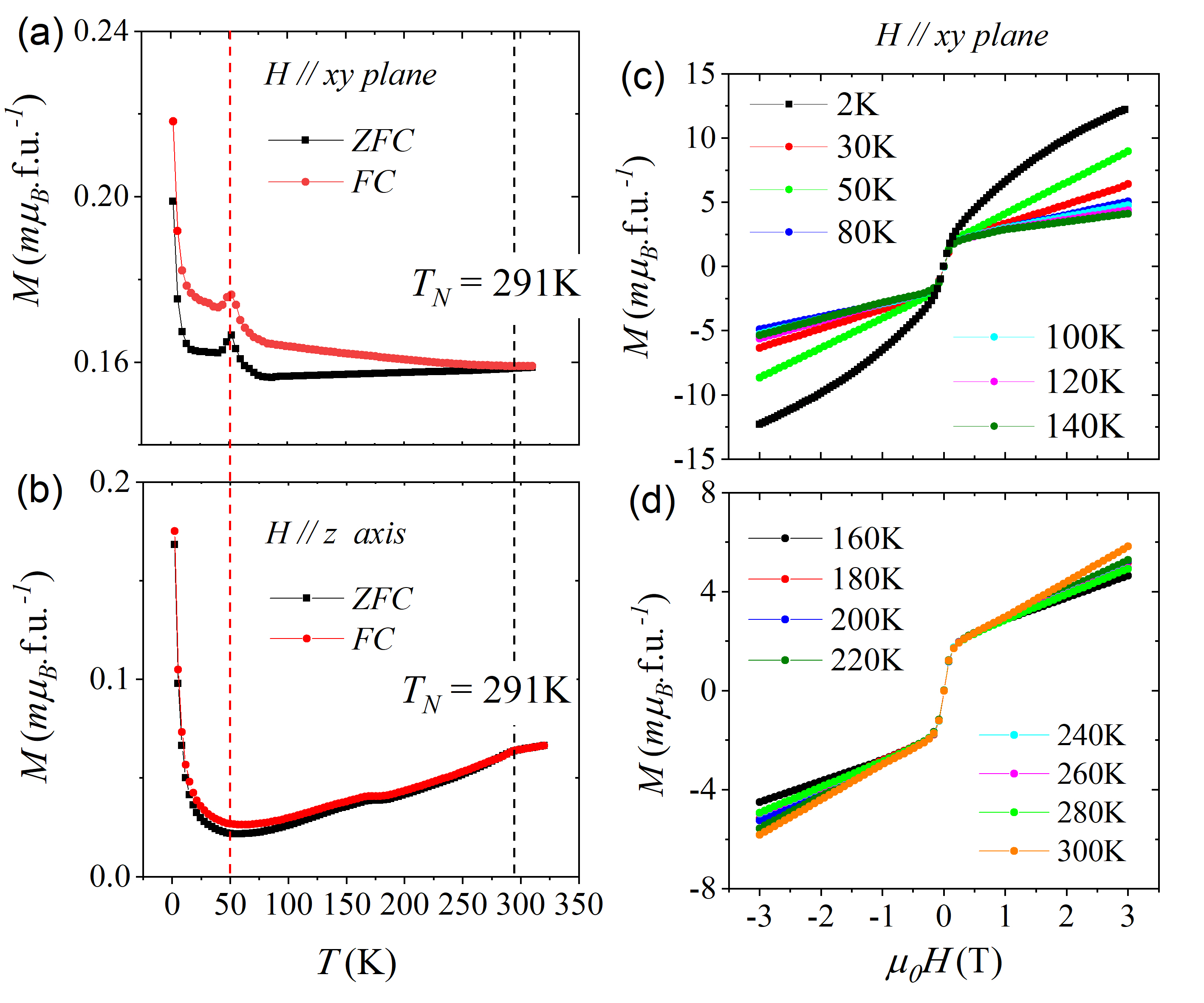}
	\textcolor{black}{\caption{ (a),(b) Temperature dependence of Zero-field-cooled (ZFC) magnetization and field-cooled (FC) magnetization  $M$ in the in-plane ($H$//$xy$) and out-plane ($H$//$z$) with 100 Oe. At 50K, the response is anisotropic. When the field is parallel to the $xy$ plane, there is a peak, but is absent in the perpendicular orientation ($H//z$).
			(c) The temperature dependence of magnetization of YbMnBi$_2$ reported in Ref.\cite{pan2022giantsm}, in which the measurement field is 100 Oe.
			(d), (e) The field dependence of the magnetization at various temperatures when the field is parallel to the $xy$ plane.}
		\label{fig:S4}}
\end{figure*}

\begin{figure*}[ht]
	\setlength {\abovecaptionskip} {0.2cm}
	\setlength {\belowcaptionskip} {-0.2cm}
	\includegraphics[width=15cm]{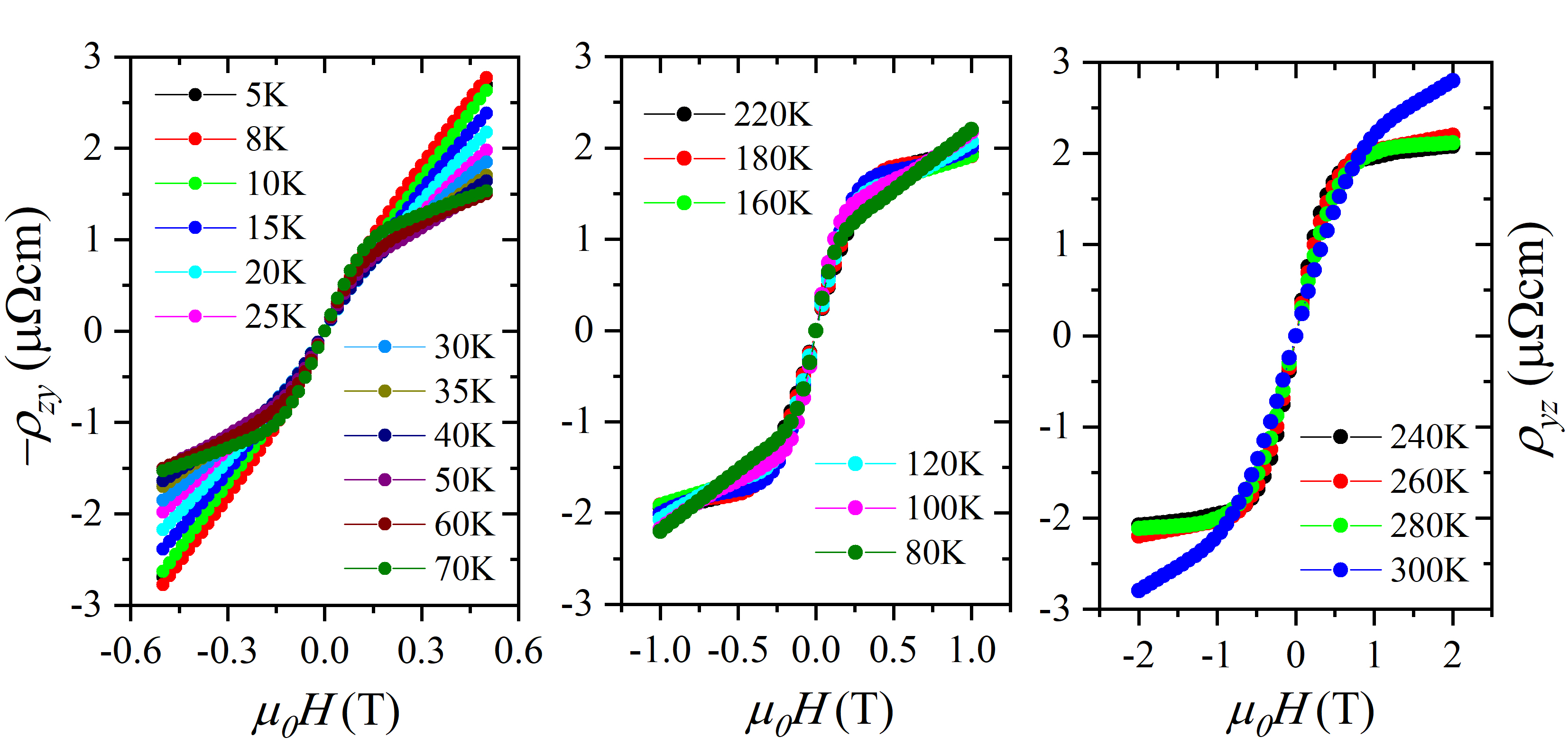}
	\textcolor{black}{\caption{  The magnetic field dependence of the Hall effect at different temperatures. For $zy$ and $yz$ configurations, the Hall resistivity is identical: $\rho_{zy}$($H$) =  $\rho_{yz}$(-$H$). }
		\label{fig:S5}}
\end{figure*}

\begin{figure*}[ht]
	\setlength {\abovecaptionskip} {0.2cm}
	\setlength {\belowcaptionskip} {-0.2cm}
	\includegraphics[width=16cm]{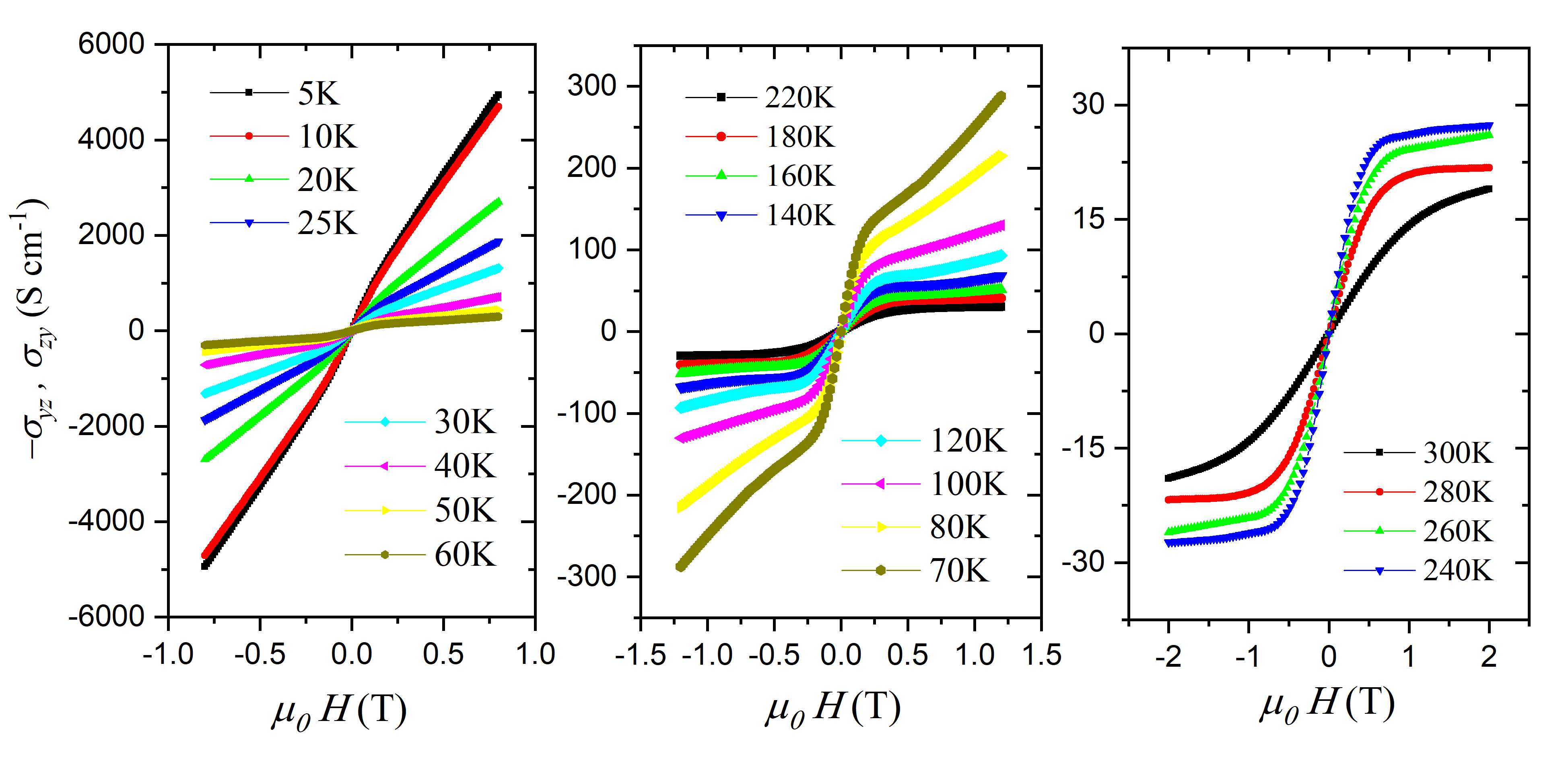}
	\textcolor{black}{\caption{The field dependence of Hall conductivity at different temperatures.  The Hall conductivity is identical, in the two configurations :$\sigma_{zy}$($H$) =  $\sigma_{yz}$(-$H$). }
		\label{fig:S6}}
\end{figure*}

\begin{figure*}[ht]
	\setlength {\abovecaptionskip} {0.2cm}
	\setlength {\belowcaptionskip} {-0.2cm}
	\includegraphics[width=17cm]{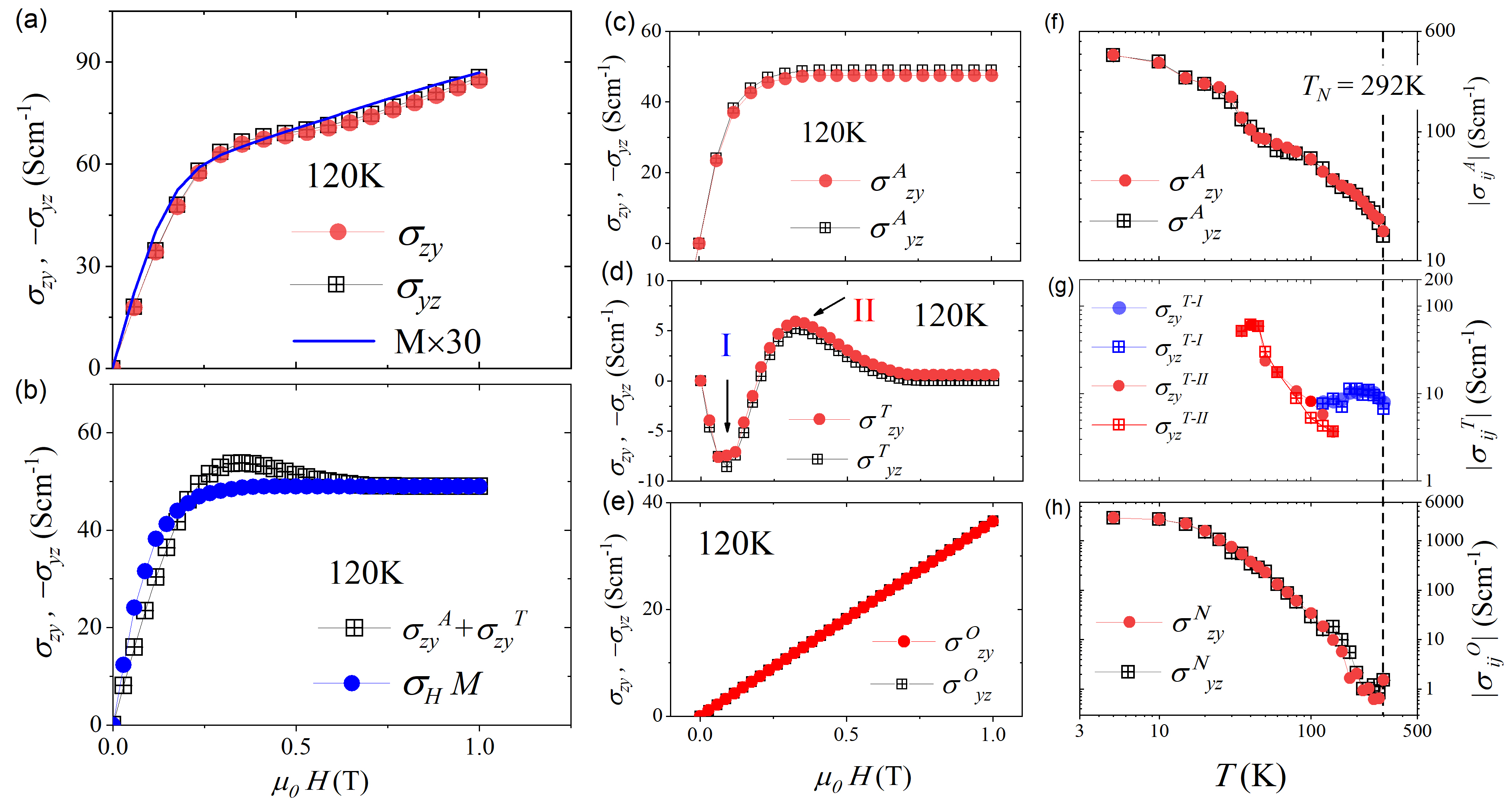}
	\textcolor{black}{\caption{Hall conductivity analyzed using equation S2. (a) Hall conductivity in the $zy$ and $yz$ configuration compared with the magnetization at the 120 K. Note the peak around 0.3 T. (b) The Hall conductivity minus with its ordinary component subtracted compared with the product of magnetization and the ordinary Hall effect. The maximum near 0.3 T is attributed to the topological Hall effect. The anomalous, topological, and ordinary Hall conductivities are displayed in (c), (d), and (e). The temperature dependence of these three components are shown in (f), (g), and (h). }
		\label{fig:S7}}
\end{figure*}

\begin{figure*}[ht]
	\setlength {\abovecaptionskip} {0.2cm}
	\setlength {\belowcaptionskip} {-0.2cm}
	\includegraphics[width=17cm]{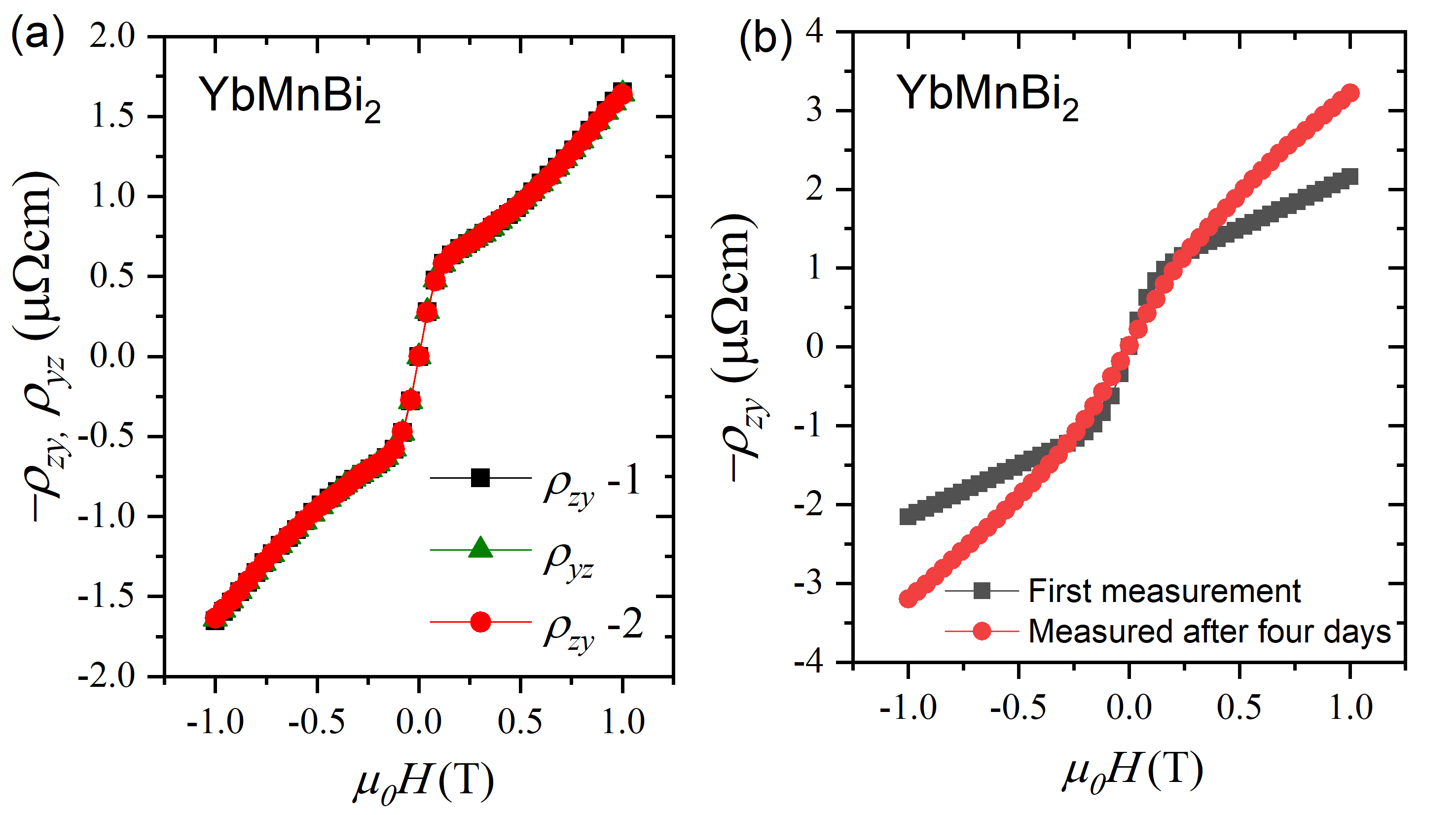}
	\textcolor{black}{\caption{ Oxidation of YbMnBi$_2$. (a) The properties do not change when the sample undergoes a circulation measurement ($\rho_{zy}\rightarrow\rho_{yz}\rightarrow\rho_{zy}$) under high vacuum. (b) Hall resistivity is
			different at identical temperatures in the same sample when it is stored in a poor vacuum environment for four days.}
		\label{fig:S8}}
\end{figure*}

\begin{figure*}[ht]
	\setlength {\abovecaptionskip} {0.2cm}
	\setlength {\belowcaptionskip} {-0.2cm}
	\includegraphics[width=16cm]{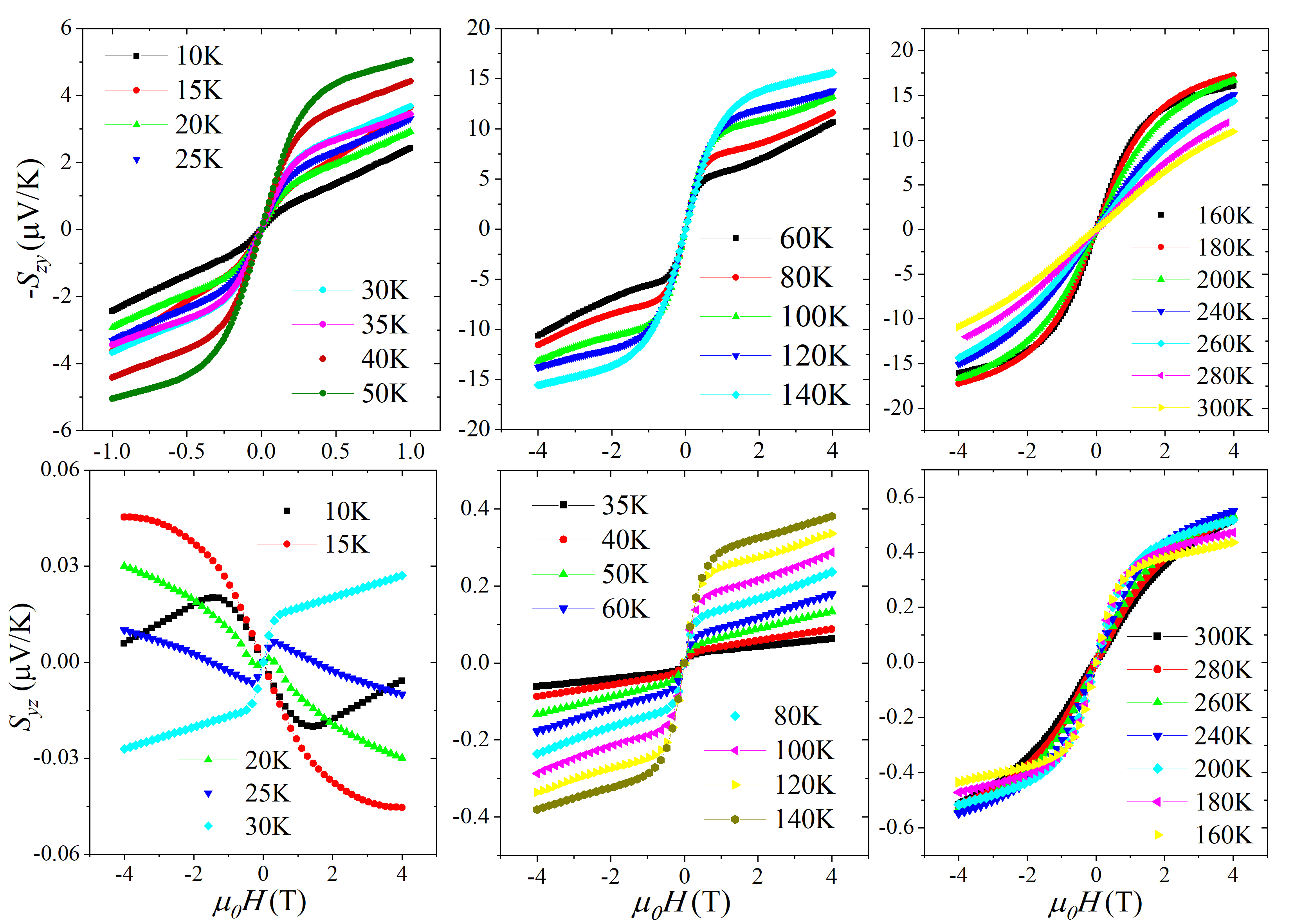}
	\textcolor{black}{\caption{ The magnetic field dependence Nernst effect at different temperatures in $zy$ and $yz$ configurations. In $zy$ and $yz$ configurations, Nernst signal is dramatically different: $\alpha_{zy}$($H$)   $\neq\alpha_{yz}$(-$H$).}
		\label{fig:S9}}
\end{figure*}

\begin{figure*}[ht]
	\setlength {\abovecaptionskip} {0.2cm}
	\setlength {\belowcaptionskip} {-0.2cm}
	\includegraphics[width=16cm]{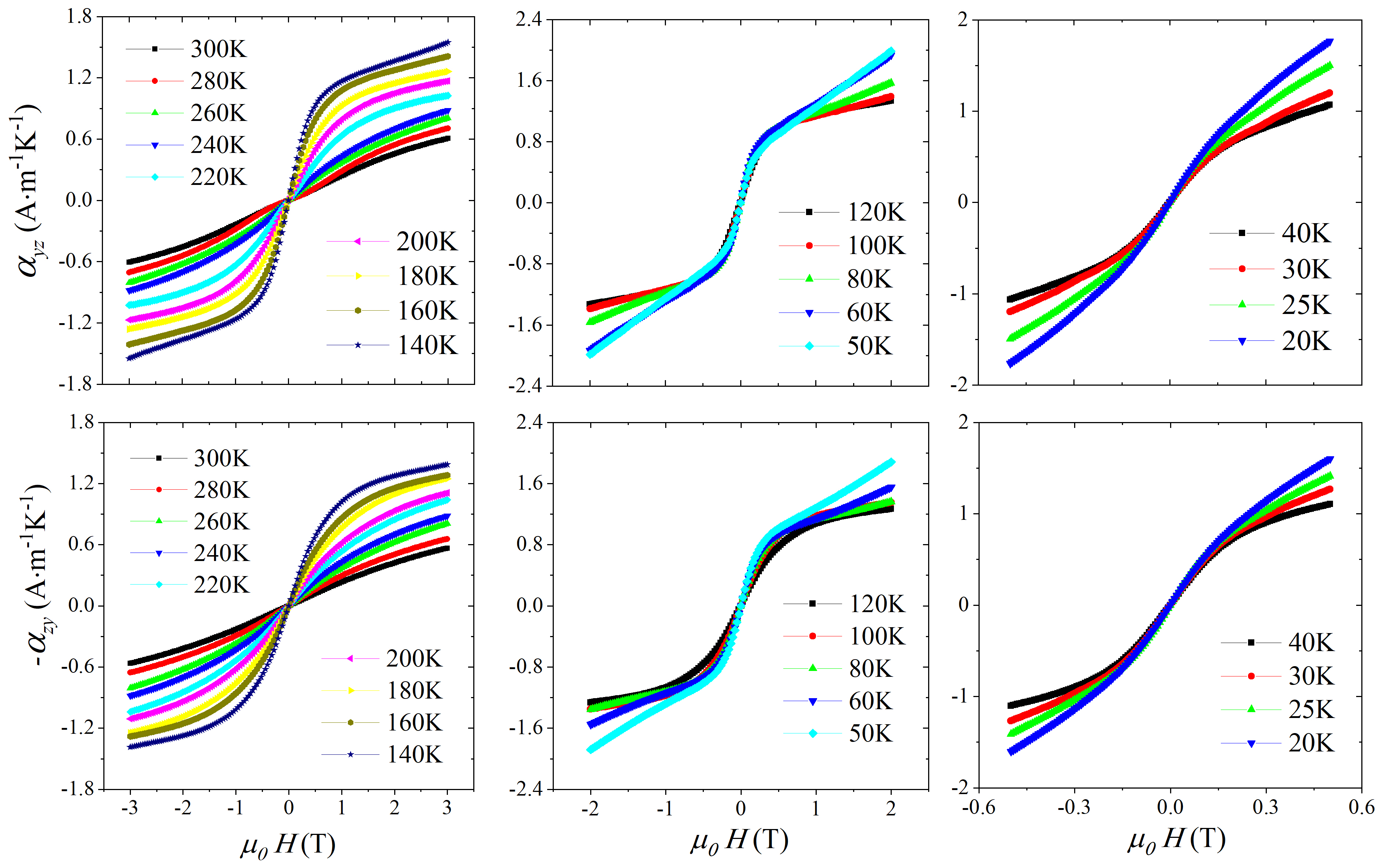}
	\textcolor{black}{\caption{The field dependence of Nernst conductivity at different temperatures. In $zy$ and $yz$ configurations, Nernst conductivity is identical: $\alpha_{zy}$($H$) =  $\alpha_{yz}$(-$H$).}
		\label{fig:S10}}
\end{figure*}

\begin{figure*}[ht]
	\setlength {\abovecaptionskip} {0.2cm}
	\setlength {\belowcaptionskip} {-0.2cm}
	\includegraphics[width=17cm]{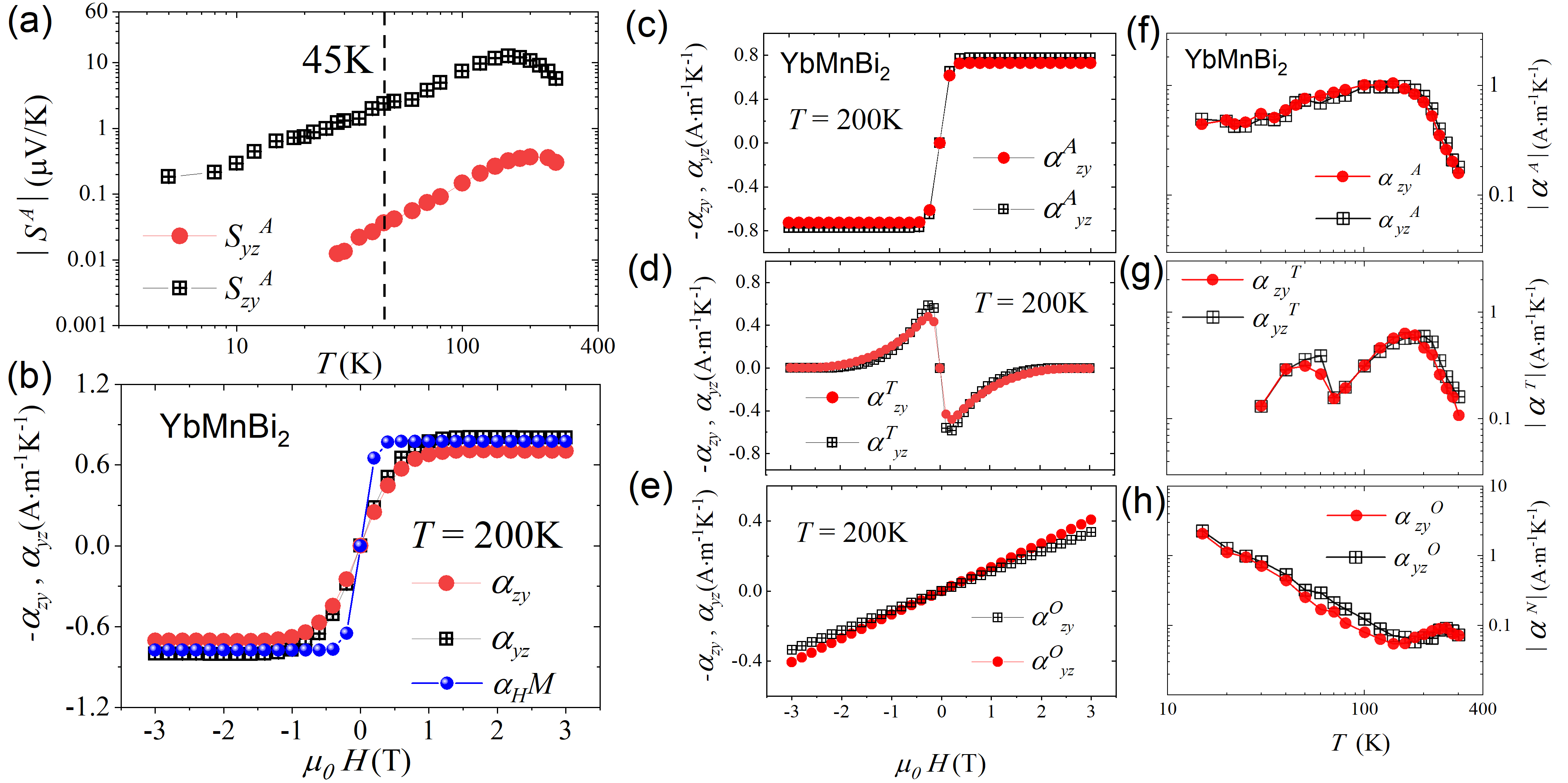}
	\textcolor{black}{\caption{ (a) The anomalous Nernst thermopower in the $zy$ and $yz$ configurations. (b) The Nernst conductivity in the $zy$ and $yz$ configuration compared with the magnetization at 200 K. The anomalous, topological, and ordinary Nernst conductivities are displayed in (c), (d), and (e). The temperature dependence of the three components (anomalous, topological, and ordinary) is shown in (f), (g), and (h).}
		\label{fig:S11}}
\end{figure*}

\begin{figure*}[ht]
	\setlength {\abovecaptionskip} {0.2cm}
	\setlength {\belowcaptionskip} {-0.2cm}
	\includegraphics[width=14cm]{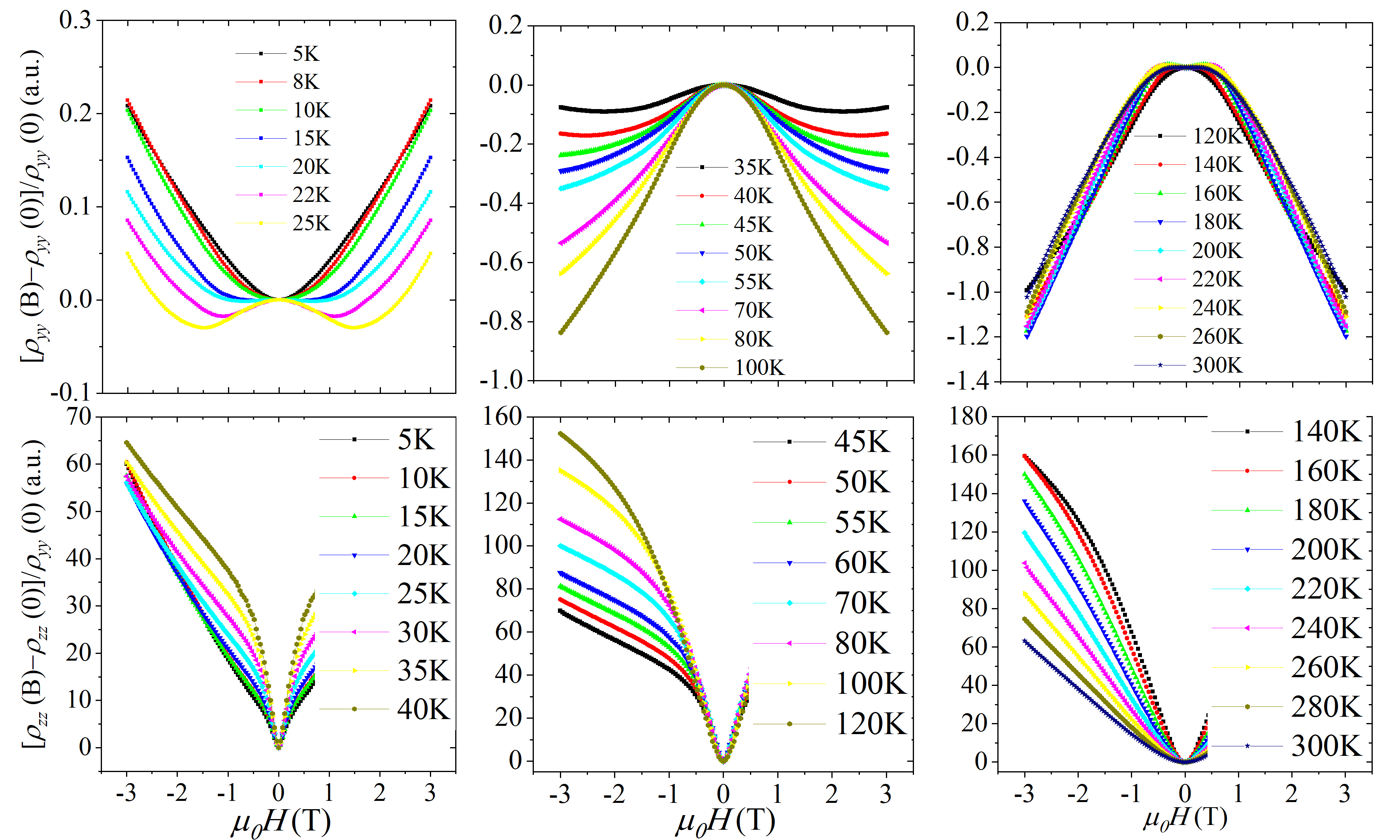}
	\textcolor{black}{\caption{The field dependence of resistivity at various temperatures for the two configurations. (Upper panels). $\rho_{yy}$ ( $H$//$x$, with the current along the $y$ axis, $I$//$y$). Negative magnetoresistance is observed when the temperature is above 15 K . (Lower panels). When the applied current is along the $z$ axis and $H$//$x$, magnetoresistance $\rho_{zz}$ is always positive. }
		\label{fig:S12}}
\end{figure*}

\begin{figure*}[ht]
	\setlength {\abovecaptionskip} {0.2cm}
	\setlength {\belowcaptionskip} {-0.2cm}
	\includegraphics[width=16cm]{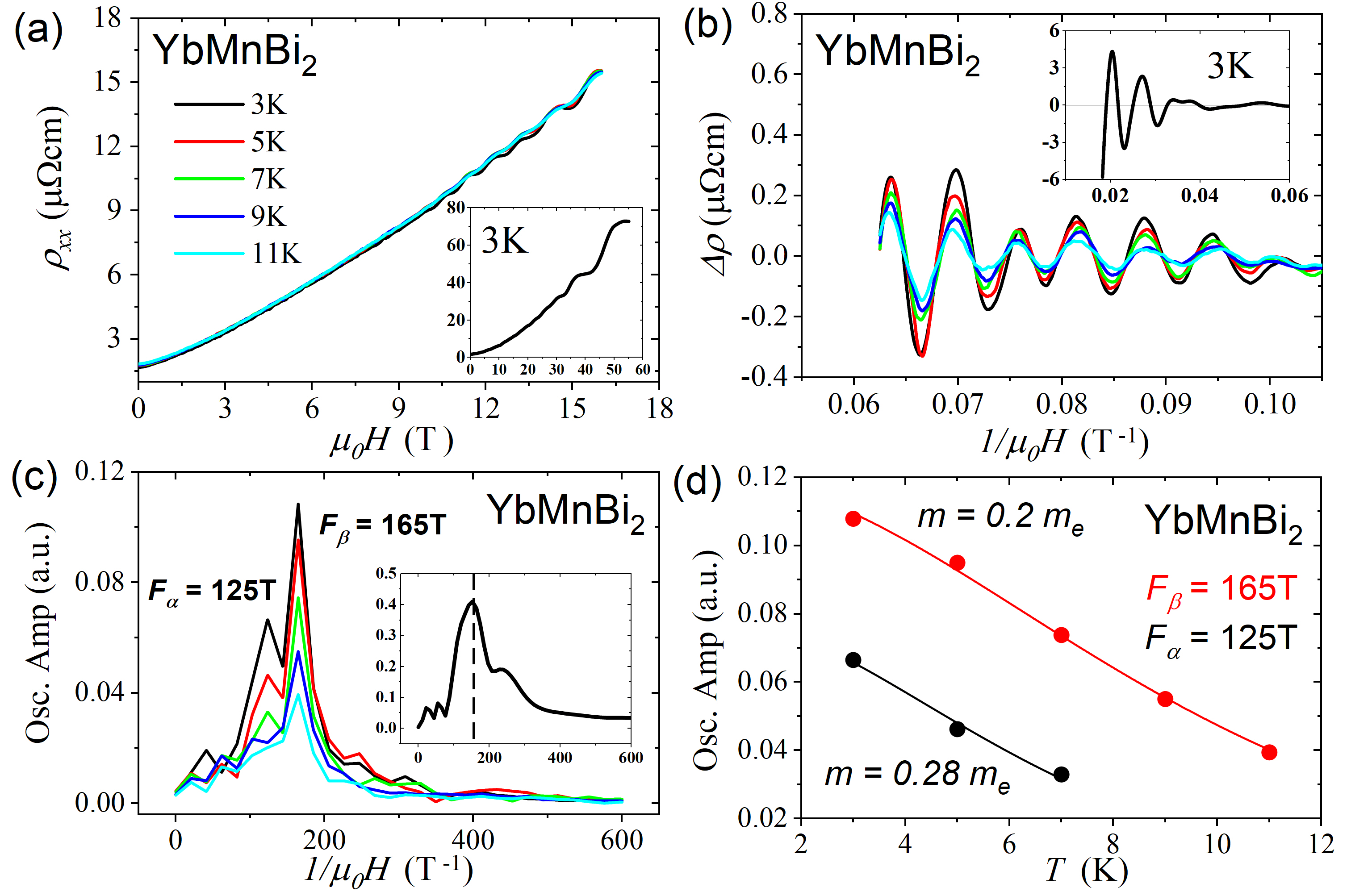}
	\textcolor{black}{\caption{ (a) The field dependence of resistivity, $\rho_{xx}$ ( B//$z$, when the current is along the $x$ axial, $I$//$x$). The obvious quantum oscillation can be observed at 3 K when the field is above 30 T up to 55 T as shown in the insert. (b) Oscillations after the background subtraction. The insert shows the oscillation component of magnetoresistivity obtained by the pulsed high magnetic field at 3 K. (c) The FFT spectrum of the oscillations. Also, the insert displays the FFT spectrum observed in the pulsed high magnetic field. (d) The temperature dependence of the oscillating amplitude of magnetoresistivity. The solid line is the fitting curve with the Lifshitz-Kosevich model. }
		\label{fig:S13}}
\end{figure*}

\begin{figure*}[ht]
	\setlength {\abovecaptionskip} {0.4cm}
	\setlength {\belowcaptionskip} {-0.1cm}
	\includegraphics[width=17cm]{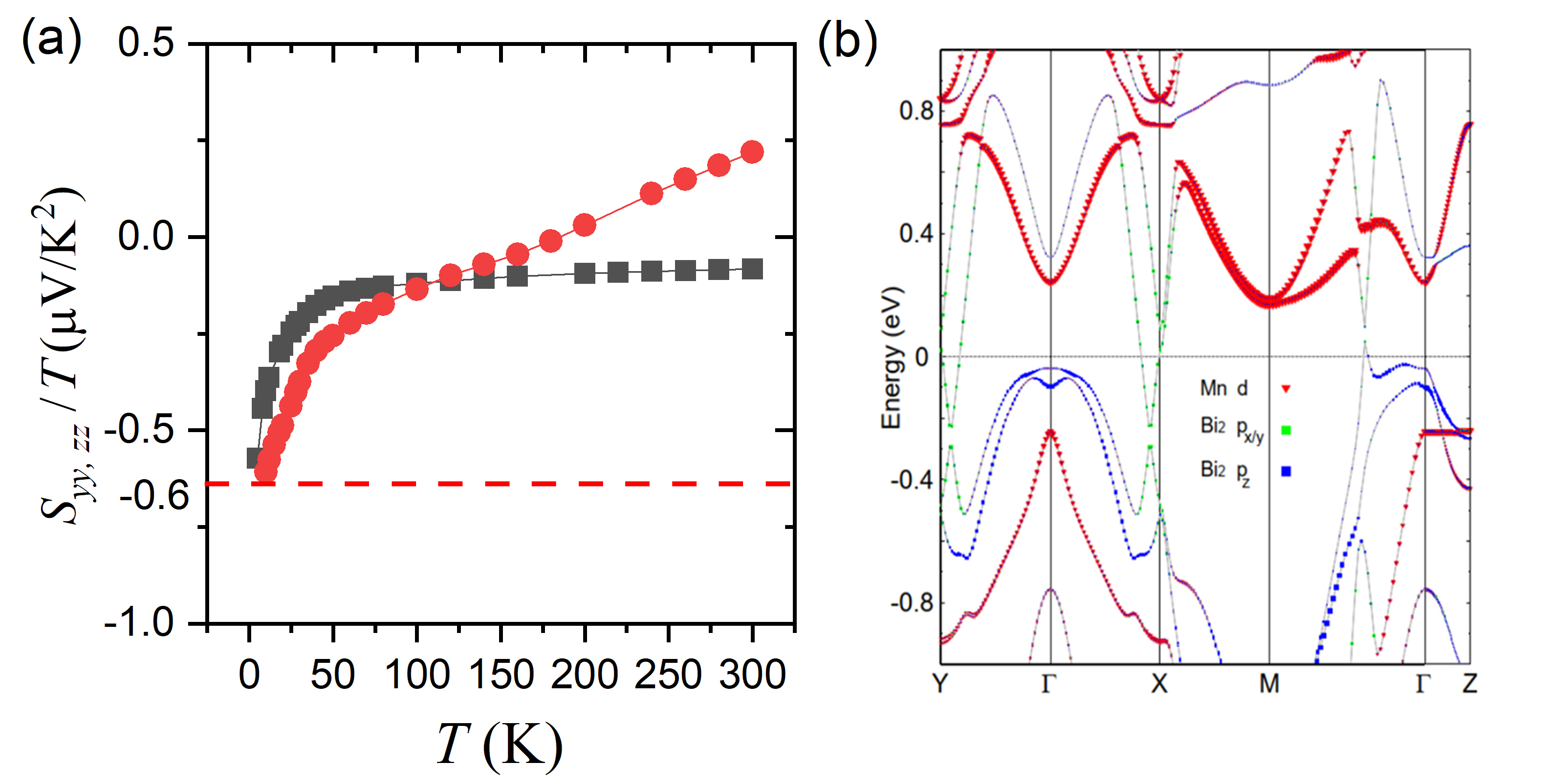}
	\textcolor{black} {\caption{  (a) The temperature dependence of the $S_{ii}/T$. $S_{ii}/T$ becomes isotropic at low temperatures. Its rough amplitude of  $\approx 0.6 \mu$V/K$^2$  is consistent with a Fermi energy of 76 meV  (b) The Fermi level obtained by DFT calculation\cite{pan2022giantsm}. Fermi energy is higher for electrons pockets than for hole pockets.}}
	\label{fig:S14} 
\end{figure*}

\begin{figure*}[ht]
	\setlength {\abovecaptionskip} {0.5cm}
	\setlength {\belowcaptionskip} {-0.2 cm}
	\includegraphics[width=17cm]{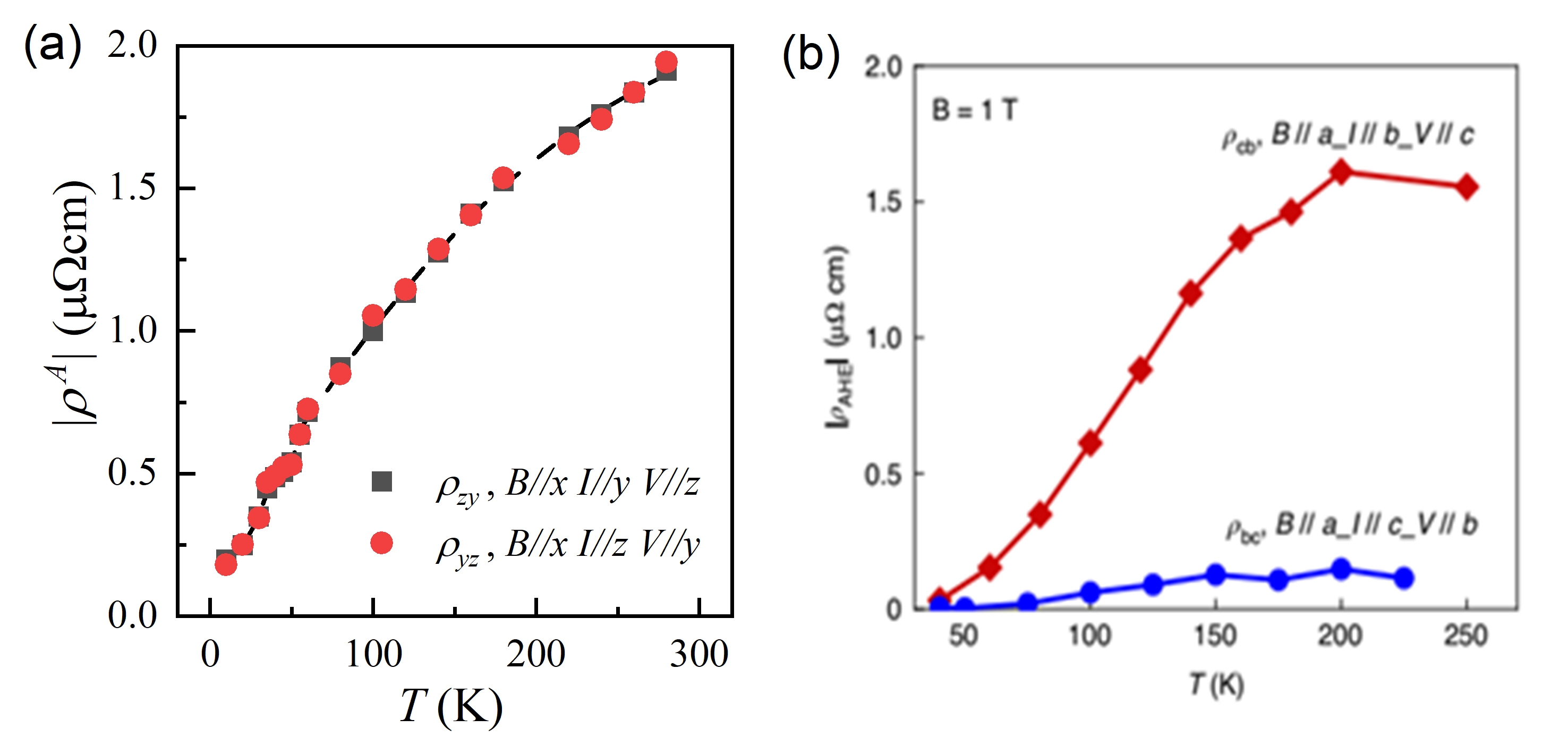}
	\textcolor{black} {\caption{(a) the temperature dependence of anomalous Hall effect in our work. (b) the temperature dependence of anomalous Hall effect in Ref.\cite{pan2022giantsm}}}
	\label{fig:S15} 
\end{figure*}

\begin{figure}[ht]
	\setlength {\abovecaptionskip} {-0.3cm}
	\setlength {\belowcaptionskip} {-0.2 cm}
	\includegraphics[width=9cm]{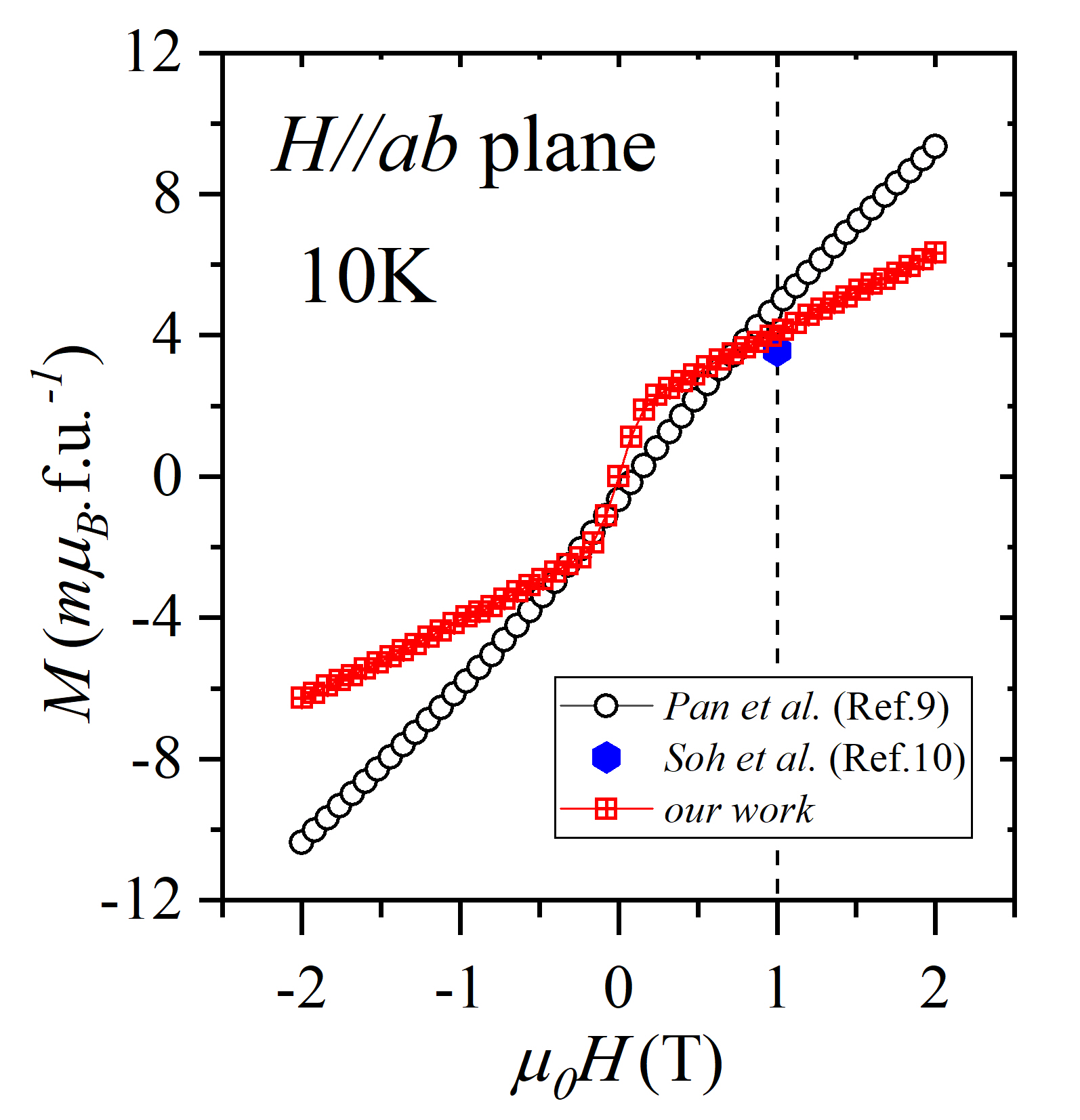}
	\textcolor{black} {\caption{The field dependence of magnetization at 10K\cite{pan2022giantsm,soh2019magneticsm}. }}
	\label{fig:S16} 
\end{figure}


\providecommand{\noopsort}[1]{}\providecommand{\singleletter}[1]{#1}%

\end{document}